\documentclass[fleqn,usenatbib]{mnras}

\usepackage{newtxtext,newtxmath}

\usepackage[T1]{fontenc}
\usepackage{ae,aecompl}

\usepackage{graphicx}	
\usepackage{amsmath}	
\usepackage{amssymb}	
\usepackage{lineno}
\usepackage{aas_macros}

\newcommand{\Msun}{M$_{\sun}$}
\newcommand{\Rsun}{R$_{\sun}$}
\graphicspath{{./}{figures/}}
\newcommand{\nifs}{\ensuremath{^{56}}Ni}
\newcommand{\cofs}{\ensuremath{^{56}}Co}
\newcommand{\nodata}{\centering--}

\usepackage{eso-pic}

\title[DES16C3cje: A low-luminosity, long-lived supernova]{DES16C3cje: A low-luminosity, long-lived supernova}

\author[C.~P. Guti\'errez et al.]{
\parbox{\textwidth}{
\Large
C.~P.~Guti\'errez,$^{1}$\thanks{E-mail: C.P.Gutierrez-Avendano@soton.ac.uk}
M.~Sullivan,$^{1}$
L.~Martinez,$^{2,3}$
M.~C.~Bersten,$^{2,3,4}$
C.~Inserra,$^{5}$
M.~Smith,$^{1}$
J.~P.~Anderson,$^{6}$
Y.-C.~Pan,$^{7}$ 
A.~Pastorello,$^{8}$
L. Galbany,$^{9}$
P.~Nugent,$^{10}$
C.~R.~Angus,$^{11}$
C.~Barbarino,$^{12}$
D.~Carollo,$^{13}$
T.-W.~Chen,$^{14}$
T.~M.~Davis,$^{15}$
M.~Della Valle,$^{16,17}$
R.~J.~Foley,$^{18}$
M.~Fraser,$^{19}$
C.~Frohmaier,$^{20}$
S. Gonz\'alez-Gait\'an,$^{21}$
M.~Gromadzki,$^{22}$
E.~Kankare,$^{23}$
R.~Kokotanekova,$^{17}$
J.~Kollmeier,$^{24}$
G.~F.~Lewis,$^{25}$
M.~R.~Magee,$^{26}$
K.~Maguire,$^{26}$
A.~M\"oller,$^{27}$
N.~Morrell,$^{28}$
M.~Nicholl,$^{29,30}$
M.~Pursiainen,$^{1}$
J.~Sollerman,$^{12}$
N.~E.~Sommer,$^{31}$
E.~Swann,$^{20}$
B.~E.~Tucker,$^{31}$
P.~Wiseman,$^{1}$
M.~Aguena,$^{32,33}$
S.~Allam,$^{34}$
S.~Avila,$^{35}$
E.~Bertin,$^{36,37}$
D.~Brooks,$^{38}$
E.~Buckley-Geer,$^{34}$
D.~L.~Burke,$^{39,40}$
A.~Carnero~Rosell,$^{41,33}$
M.~Carrasco~Kind,$^{42,43}$
J.~Carretero,$^{44}$
M.~Costanzi,$^{45,46}$
L.~N.~da Costa,$^{33,47}$
J.~De~Vicente,$^{41}$
S.~Desai,$^{48}$
H.~T.~Diehl,$^{34}$
P.~Doel,$^{38}$
T.~F.~Eifler,$^{49,50}$
B.~Flaugher,$^{34}$
P.~Fosalba,$^{51,52}$
J.~Frieman,$^{34,53}$
J.~Garc\'ia-Bellido,$^{35}$
D.~W.~Gerdes,$^{54,55}$
D.~Gruen,$^{56,39,40}$
R.~A.~Gruendl,$^{42,43}$
J.~Gschwend,$^{33,47}$
G.~Gutierrez,$^{34}$
S.~R.~Hinton,$^{15}$
D.~L.~Hollowood,$^{18}$
K.~Honscheid,$^{57,58}$
D.~J.~James,$^{59}$
K.~Kuehn,$^{60,61}$
N.~Kuropatkin,$^{34}$
O.~Lahav,$^{38}$
M.~Lima,$^{32,33}$
M.~A.~G.~Maia,$^{33,47}$
M.~March,$^{62}$
F.~Menanteau,$^{42,43}$
R.~Miquel,$^{63,44}$
E.~Morganson,$^{43}$
A.~Palmese,$^{34,53}$
F.~Paz-Chinch\'{o}n,$^{42,43}$
A.~A.~Plazas,$^{64}$
M.~Sako,$^{62}$
E.~Sanchez,$^{41}$
V.~Scarpine,$^{34}$
M.~Schubnell,$^{55}$
S.~Serrano,$^{51,52}$
I.~Sevilla-Noarbe,$^{41}$
M.~Soares-Santos,$^{65}$
E.~Suchyta,$^{66}$
M.~E.~C.~Swanson,$^{43}$
G.~Tarle,$^{55}$
D.~Thomas,$^{20}$
T.~N.~Varga,$^{14,67}$
A.~R.~Walker,$^{68}$
R.~Wilkinson$^{69}$
\begin{center} (DES Collaboration) \end{center}
}
\vspace{0.4cm}
\\
\parbox{\textwidth}{Affiliations are listed at the end of the paper}
}

\date{Accepted XXX. Received YYY; in original form ZZZ}

\pubyear{2019}

\begin{document}
\label{firstpage}
\pagerange{\pageref{firstpage}--\pageref{lastpage}}
\maketitle

\begin{abstract}
We present DES16C3cje, a low-luminosity, long-lived type II supernova (SN~II) at redshift 0.0618, detected by the Dark Energy Survey (DES). DES16C3cje is a unique SN. The spectra are characterized by extremely narrow photospheric lines corresponding to very low expansion velocities of $\lesssim1500$\,km\,s$^{-1}$, and the light curve shows an initial peak that fades after 50 days before slowly rebrightening over a further 100 days to reach an absolute brightness of M$_r\sim -15.5$\,mag. The decline rate of the late-time light curve is then slower than that expected from the powering by radioactive decay of \cofs, but is comparable to that expected from accretion power. Comparing the bolometric light curve with hydrodynamical models, we find that DES16C3cje can be explained by either i) a low explosion energy (0.11\,foe) and relatively large \nifs\ production of 0.075\,\Msun\ from a $\sim15$\,\Msun\ red supergiant progenitor typical of other SNe~II, or ii) a relatively compact $\sim40$\,\Msun\ star, explosion energy of 1\,foe, and 0.08\,\Msun\ of \nifs. Both scenarios require additional energy input to explain the late-time light curve, which is consistent with fallback accretion at a rate of $\sim0.5\times{10^{-8}}$\,M$_{\odot}$\,s$^{-1}$.
\end{abstract}




\section{Introduction}
\label{intro}

Recent wide-field sky surveys have revealed a significant diversity in the observed properties of supernovae (SNe). These events have covered a wide range of observed characteristics: transients with extremely bright luminosities \citep[e.g., superluminous SNe,][]{GalYam12}; transients with a rapid temporal evolution spanning a range of luminosities \citep[e.g.,][]{Perets10,Kasliwal12, Drout14, Pursiainen18}, and a heterogeneous population of transients with a slow temporal evolution \citep[e.g.,][]{Taddia16a,Arcavi17,Terreran17}. These new SN discoveries have in turn created new challenges for the SN field, particularly concerning the SN progenitor and the physics of the explosion.

In the canonical picture of a core-collapse SN, the explosion releases $\sim10^{51}$\,erg of energy (1\,foe), and a fraction of the progenitor's material is burned into various intermediate-mass and iron-peak elements. 
The early emission from SNe, defined as the cooling phase, is powered by the release of shock deposited energy, while the power source from the peak to late-phases is provided by the decay of \nifs\ into \cofs\ and subsequently $^{56}$Fe. In slow- and fast-declining hydrogen-rich SNe (historical SNe~IIP and SNe~IIL, respectively), the cooling phase is followed by a hydrogen recombination phase, where the luminosity evolves more slowly until it becomes dominated by the energy released during the decay of radioactive material. 
However, some core-collapse SNe have larger luminosities, which typically require an additional source of energy to explain them \citep[see review, and references therein, of][]{Moriya18}. Pair-Instability SNe (PISNe; e.g. \citealt[][]{Heger02,GalYam09a}), magnetars \citep[e.g.][]{Kasen10a,Bersten16}, accretion power \citep[e.g.][]{Moriya10, Dexter13}, and pulsational pair-instability (PPI; e.g. \citealt[][]{Woosley07,Woosley17}) have all been proposed as a source of additional energy, but as yet there is no clear consensus about the relative importance of each source nor associations to specific transients.

Recently, two peculiar type II SNe (SNe~II) have been studied in detail: iPTF14hls \citep{Arcavi17,Sollerman18} and OGLE-2014-SN-073 \citep{Terreran17}. iPTF14hls is a SN with very little spectral evolution over $\sim600$ days, and with a light curve that shows multiple re-brightening events. OGLE-2014-SN-073 is a very bright SN with an unusually broad light curve, combined with high ejecta velocities in its spectra. Both objects exploded in low-luminosity galaxies and require an extra source of power (beyond shock energy and radioactivity) to explain their unusual evolution. 

Popular scenarios invoked to explain the peculiar behaviour of these two transients are a magnetar \citep{Dessart18,Orellana18,Woosley18}, PISNe \citep{Woosley18}, circumstellar interaction \citep{Andrews18,Woosley18} and fallback accretion \citep{Arcavi17,Moriya18a,Wang18}. \citet{Moriya18a} found the latter scenario can reproduce the shape of the light curve, luminosity and photospheric velocities of OGLE-2014-SN-073, while \citet{Arcavi17} and \citet{Wang18} proposed that iPTF14hls may be powered by intermittent fallback accretion. The idea of fallback in SNe was introduced by \citet{Colgate71}, and has been broadly studied to determine its effects on the central remnant \citep[e.g.][]{Chevalier89, Woosley95, Fryer99}, and on SN light curves \citep[e.g.][]{Fryer09, Moriya10, Dexter13}. \citet{Dexter13} showed that the accretion power may be relevant to explain peculiar and rare SNe.

In this paper, we present the photometry and spectra of DES16C3cje, an unusual SN II discovered by the Dark Energy Survey Supernova Program \citep[DES-SN;][]{Bernstein2012}. We discuss its peculiar characteristics and examine the late-time light curve under  the fallback scenario. In Section~\ref{sec:obs} we describe our observations of DES16C3cje and measurements. We analyse the spectral and photometric properties and compare them with other similar events in Section~\ref{sec:results}, and then discuss the progenitor scenarios that could explain the event in Section~\ref{sec:modelling}. We discuss and conclude in Section~\ref{sec:summary}. Throughout, we assume a flat $\Lambda$CDM universe, with a Hubble constant of $H_0=70$\,km\,s$^{-1}$\,Mpc$^{-1}$, and $\Omega_\mathrm{m}=$0.3.

\section{Observations}
\label{sec:obs}

DES16C3cje was detected by DES using the wide-field Dark Energy Camera \citep[DECam;][]{Flaugher15} instrument  in an $r$-band image taken on 2016 October 11 (JD = 2457673.3) with an apparent magnitude of $r=23.26$\,mag. The transient was located at $\alpha=03^{\mathrm{h}}28^{\mathrm{m}}35{\fs}29$, $\delta=-27^\circ09'06{\farcs}6$ (J2000.0) in a faint host galaxy ($M_r\sim-18.5$\,mag) at a redshift of 0.0616. The previous non-detection with DES was obtained on 2016 October 7 ($\mathrm{MJD} = 57667.6$), with a detection limit of $z\sim25.1$\,mag. This limit places a constraint on the explosion epoch of $\pm2.6$ days; we adopt 2016 October 9 (the intermediate epoch; $\mathrm{MJD}=57670.2\pm2.6$\,d) as the explosion date. Further information on the DES-SN difference-imaging search pipeline and machine-learning algorithms to identify transient objects can be found in \citet{Kessler15} and \citet{Goldstein2015}.

Photometric coverage of DES16C3cje was acquired by DES-SN in $griz$ filters from 2016 October until 2017 February, and then from 2017 August to 2018 February. Between 2017 February and 2017 July, additional photometric data were obtained by the extended Public European Southern Observatory (ESO) Spectroscopic Survey for Transient Objects \citep[ePESSTO;][]{Smartt15a} and other collaborators with the ESO Faint Object Spectrograph and Camera 2 \citep[EFOSC2;][]{Buzzoni84} at the 3.6m ESO New Technology Telescope (NTT), with the FOcal Reducer/low dispersion Spectrograph 2 \citep[FORS2;][]{Appenzeller98} at the ESO Very Large Telescope (VLT), with the Low Dispersion Survey Spectrograph 3 \citep[LDSS3;][]{Osip04} on the Magellan Clay 6.5-m telescope, and with the the Gamma-Ray Burst Optical/Near-Infrared Detector \citep[GROND;][]{ Greiner08}, at the 2.2-m MPG telescope at the European Southern Observatory (ESO) La Silla Observatory.

The NTT data were reduced using the PESSTO pipeline \citep{Smartt15a}, while for the FORS2 images we used the \textsc{esoreflex} pipeline \citep{Freudling13}. Reductions for data obtained with LDSS3 were performed with Image Reduction and Analysis Facility \citep[\textsc{iraf};][]{Tody86} using standard routines. 
Images from the MPG were reduced with the GROND pipeline \citep{Kruhler08}.
The DES photometric measurements were made using the pipeline discussed by \citet{Papadopoulos2015} and \citet{Smith2016}, which has also been extensively used in the literature \citep[e.g.,][and references therein]{PTFphot}. This pipeline subtracts a deep template image from each individual DES image to remove the host-galaxy light using a point-spread-function (PSF) matching routine. SN photometry is then measured from the difference image using a PSF-fitting technique.
The photometry of DES16C3cje is reported in Appendix~\ref{photo}. 

DES16C3cje was observed spectroscopically on six epochs from $+47$ to $+403$ days (throughout the paper, we give all epochs relative to the explosion epoch). These observations were obtained with four different instruments: The AAOmega spectrograph at the Anglo-Australian Telescope (AAT), 
X-SHOOTER \citep{2011A&A...536A.105V} and FORS2 at the VLT, and Gemini Multi-Object Spectrograph (GMOS-S; \citealt{Hook04}) at the Gemini Observatory. A log of the spectroscopic observations of DES16C3cje is reported in Table~\ref{tab:speclog}. Spectroscopic reductions for X-SHOOTER were performed using the \textsc{esoreflex} pipeline, FORS2 data were reduced with \textsc{iraf} using standard routines, while for GMOS-S  we used the Gemini \textsc{iraf} package, combined with \textsc{idl} routines to flux calibrate the data and remove telluric lines.  

\begin{table*}
\centering
\small
\caption{Spectroscopic observations of DES16C3cje.\label{tab:speclog}}
\label{spec}
\begin{tabular}[t]{ccccccccccc}
\hline
UT date		&	MJD    & Rest-frame phase$^{\star}$	&  Telescope    &  Range	&  Grism/Grating/	\\
		    & (days)   & (days) & + Instrument	&  (\AA)	&     Arm          \\
\hline
\hline               
20161127    & 57719.7  &  47    &  AAT+AAOmega   &  3750 -- 9000  & 580V+385R  \\
20170102    & 57755.6  &  80    &  Gemini+GMOS-S &  5700 -- 7500  & R400-G5305   \\
20170129    & 57782.0  &  105   &  VLT+XSHOOTER  &  3100 -- 10400 & UV/VIS/NIR   \\
20170221    & 57805.0  &  127   &  VLT+XSHOOTER  &  3100 -- 10400 & UV/VIS/NIR   \\
20170731    & 57965.3  &  278   &  VLT+FORS2     &  4300 -- 9500  & 300V+GG435  \\
20171116    & 58074.2  &  380   &  VLT+XSHOOTER  &  3600 -- 9600  & UV/VIS/NIR   \\
\hline
\hline
\end{tabular}
\begin{list}{}{}
\item $^{\star}$ The phase is relative to the estimated explosion date, MJD$=57670.2\pm2.6$ d. \\
\end{list}
\end{table*}

\section{Characterizing DES16C3cje}
\label{sec:results}

\subsection{Host galaxy properties}
\label{sec:host}

The host galaxy of DES16C3cje was identified as PGC3243310, a low-luminosity galaxy (M$_B^\mathrm{host}=-18.26\pm0.50$\,mag\footnote{http://leda.univ-lyon1.fr/}) at a redshift of 0.0618\footnote{Redshift obtained from the narrow emission lines of the host galaxy}. Adopting the recessional velocity corrected into the CMB frame\footnote{http://ned.ipac.caltech.edu/} (v$=18465\pm89$\,km\,s$^{-1}$), we obtain a distance of 275.95\,Mpc, which corresponds to $\mu=37.20$. The galactic reddening in the direction of PGC3243310 is $E(B-V)=0.17$ mag \citep{Schlafly11}. Due to the faintness of the galaxy and the absence of the absorption \ion{Na}{i}\,D lines in the SN spectra, we assume the host extinction negligible. 

Using a spectrum obtained by OzDES with the AAOmega at the AAT (see Sec.~\ref{sec:sspec}) and a  spectrum from the 2dF Galaxy Redshift Survey \citep{Colless03}, we estimate the integrated oxygen abundance. The lack of [\ion{N}{ii}] suggests a very low metallicity. Setting the upper limits of the flux ratio of H$\alpha/[$\ion{[N}{ii]}$\lambda6583$ and measuring the ratio of \ion{[O}{iii]}$\lambda5007/$H$\beta$, we estimate the upper limit of the metallicity. Applying the O3N2 diagnostic method from \citet{Marino13}, we obtain an oxygen abundance of $12+\log(\mathrm{O}/\mathrm{H})<8.19\pm0.02$. With the luminosity of H$\alpha$ and the equation of \citet{Kennicutt12}, we calculate the SFR to be 0.042\,\Msun\,yr$^{-1}$.

\subsection{Light curves}
\label{sec:lightcurves}

\begin{figure*}
\centering
\includegraphics[width=0.63\textwidth]{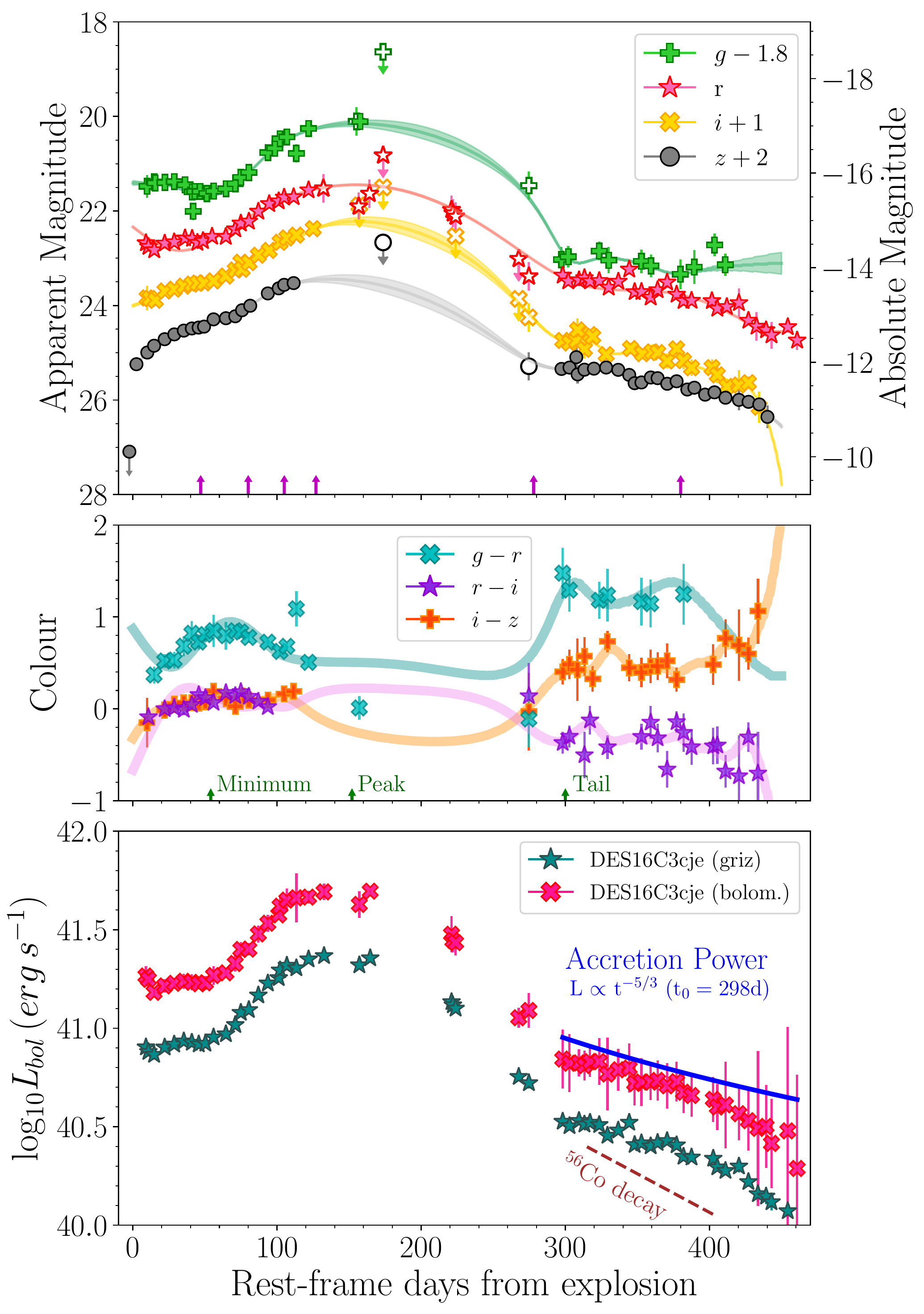}
\caption{\textbf{Upper:} $griz$ light curves of DES16C3cje. Filled symbols represent the data obtained with DES, while open stars show the data obtained with EFOSC2, LDSS3, FORS2 and GROND. Only corrections for Milky Way extinction have been made. The last non-detection is presented as a green arrow. Vertical purple arrows represent epochs of optical spectroscopy. Solid lines show the Gaussian process (GP) interpolation. \textbf{Middle:} Colour curves of DES16C3cje. Solid lines show the GP interpolation. Vertical green arrows represent epochs of minimum, peak and the beginning of the tail in the optical light curves. \textbf{Lower:} Bolometric (pink) and pseudo-bolometric (dark cyan) light curves of DES16C3cje. The dashed line shows the luminosity expected from $^{56}$Co (assuming full trapping) and the solid line the luminosity expected from accretion power.
} 
\label{fig:lc}
\end{figure*}

The unusual photometric evolution of DES16C3cje from $\sim+2$ to $+450$ days is presented in Figure~\ref{fig:lc} (top panel). The light curves show an initial increase in brightness for the first 20 days followed by a decrease, particularly in the bluer filters, as observed in some SNe~II \citep[e.g., SN 2004em, SN~2004ek;][]{Taddia16a}. In the redder bands, the luminosity increase monotonically, with a change in the slope at $\sim60$ days. After 60 days, the $g$-band increases $\sim 1.4$\,mag over 70 days versus $\sim1.0$\,mag in $riz$.

We use Gaussian processes (GPs) to interpolate the observed light curves \citep[see][for more details]{deJaeger17,Inserra18,Angus19}. The interpolation was performed with the \texttt{Python} package \texttt{GEORGE} \citep{Ambikasaran16} using the Matern 3/2 kernel. We find that DES16C3cje reaches a peak brightness of $\sim-15.75\pm0.10$\,mag at $152\pm5$ days in the $r$-band. The long rise is reminiscent of SN~1987A, but over a longer scale; this behavior has not previously been observed in a SN~II light curve. During the later phases (after $\sim300$ days), the light curves show a linear decline in $riz$ and a flat evolution in the $g$-band. The slope of the decline in the $r$-band light curve is 0.70\,mag per 100 days, smaller than that expected from the full trapping of gamma-ray photons and positrons from the decay of \cofs\ \citep[0.98\,mag per 100 days;][]{Woosley89}.

In the middle panel of Figure~\ref{fig:lc}, the colour curves are presented.  During the first 65 days (in the plateau),  DES16C3cje becomes redder, changing from $g-r=0.37$ to $g-r=0.85$. The SN then evolves to bluer colours. At late-phases ($>+300$\,days), the object has a redder colour than during the first two months, but its evolution is relatively flat. 

\subsection{Bolometric luminosity and Nickel mass}
\label{sec:bol}

Using the $griz$ photometric data, we compute the pseudo-bolometric and bolometric light curves for DES16C3cje (Figure~\ref{fig:lc}, bottom panel) following the prescriptions presented by \citet{Inserra18a}. In this method, the $griz$ bands are converted into fluxes at the effective filter wavelengths, and then corrected for the Milky Way extinction (presented in Section~\ref{sec:host}). A spectral energy distribution (SED) is then computed over the wavelengths covered and the flux under the SED is integrated assuming zero flux beyond the integration limits. Fluxes are converted to luminosities using the adopted distance (275.95 Mpc). We determined the points on the pseudo-bolometric light curves at epochs when $griz$ were available simultaneously. Magnitudes from the missing bands were generally estimated by interpolating or 
extrapolating the light curves using low-order polynomials (n$\leq$3) and assuming constant colours from nearest epochs. Therefore, we obtain a peak luminosity of $L_\mathrm{bol}=(4.96\pm0.10)\times 10^{41}$\,erg\,s$^{-1}$, and L$_{griz}=(2.33\pm0.08)\times 10^{41}$\,erg\,s$^{-1}$. 

As expected based on the photometric data, the bolometric light curves decline slowly at late phases. This decline rate is slower than the radioactive decay of \cofs, but comparable to that expected from accretion power. Although the light curve tail does not follow the \cofs\ decay, we can still use the luminosity at late times to estimate an upper limit to the \nifs\ mass. Comparing the bolometric light curve of DES16C3cje to that of SN~1987A, we estimate the \nifs\ mass, $M(^{56}\mathrm{Ni})_\mathrm{16cje}$, as follows: 
\begin{equation}
 M(^{56}\mathrm{Ni})_\mathrm{16cje} \approx  M(^{56}\mathrm{Ni})_\mathrm{87A}\times \frac{L_\mathrm{16cje}}{L_\mathrm{87A}}\,M_{\sun},
\end{equation}
where M($^{56}$Ni)$_\mathrm{87A}=0.075\pm0.005$\,M$_{\sun}$ is the \nifs\ mass synthesised by SN~1987A \citep{Arnett96} and $L_\mathrm{87A}$ is the bolometric luminosity at a comparable epoch. This comparison gives $M(^{56}\mathrm{Ni})_\mathrm{16cje}\approx0.068$\,M$_{\sun}$, a comparatively large value for typical SN~II, but within the range of SN\,1987A-like objects \citep{Muller17,Anderson19}.

\subsection{Spectral evolution}
\label{sec:sspec}

\begin{figure*}
\centering
\includegraphics[width=18cm]{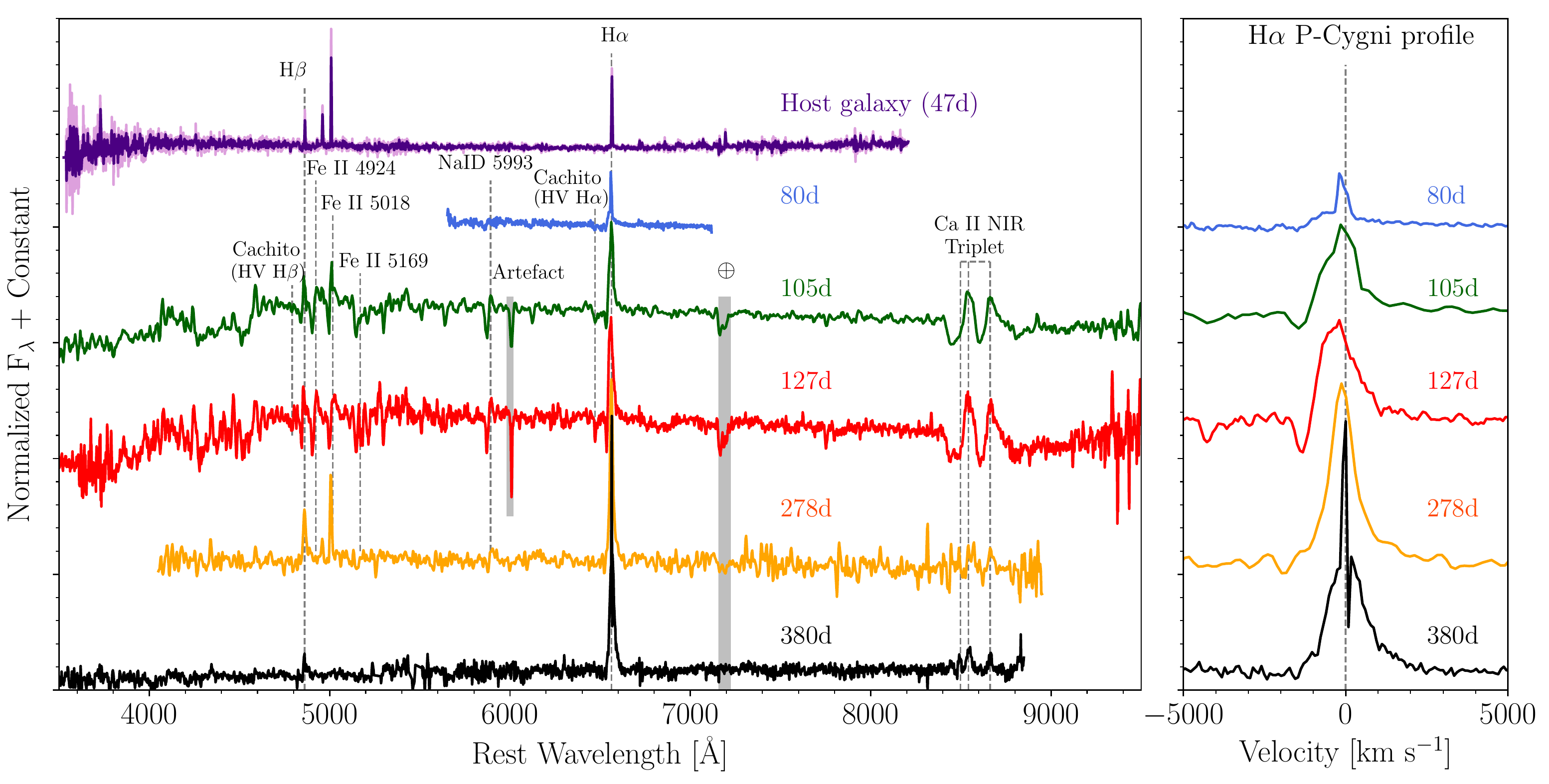}
\caption{\textbf{Left:} Optical spectra of DES16C3cje from 80 to 380 days after explosion. Each spectrum has been corrected for Milky Way reddening and shifted by an arbitrary amount for presentation. The phases are labelled on the right. \textbf{Right:} Zoom around the H$\alpha$ P-Cygni profile in velocity space. }
\label{spectra}
\end{figure*}

In Figure~\ref{spectra}, we present the optical spectra obtained for DES16C3cje between $+47$\,d and $+380$\,d. At 47\,d, the spectrum is completely dominated by the emission lines from the host galaxy, with no traces of the SN. From 80\,d, the spectra show that DES16C3cje is a SN~II with very narrow photospheric lines. At 80\,d and 127\,d, DES16C3cje presents characteristic P-Cygni profiles of H$\alpha$, H$\beta$, \ion{Fe}{ii} $\lambda4924$, \ion{Fe}{ii} $\lambda5018$, \ion{Fe}{ii} $\lambda5169$, \ion{Na}{i}\,D $\lambda5893$ and the \ion{Ca}{ii} near-IR triplet, together with a lack of \ion{Sc}{ii} and \ion{Ba}{ii} lines. The \lq Cachito\rq\ feature, related to high velocity (HV) spectra components \citep{Gutierrez17a}, are also visible at these epochs, suggesting an interaction between the SN ejecta and circumstellar material (CSM). The later spectra are dominated by H$\alpha$, with a weak contribution of the \ion{Ca}{ii} near-IR triplet in emission. There is no evidence of forbidden lines (e.g., \ion{[O}{i]} $\lambda\lambda$6300, 6363, \ion{[Fe}{ii]} $\lambda7155$ and \ion{[Ca}{ii]} $\lambda$7291, 7323), which are typical of core-collapse SNe at late phases. The lack of these lines could suggest either a high density associated with a large mass and low-velocity or an interaction between the SN ejecta and the CSM (Sec.~\ref{sec:summary}).

DES16C3cje shows a complex H$\alpha$ P-cygni profile (Figure~\ref{spectra}, right panel). At early times (spectra between 80\,d and 127\,d), the absorption component increases in strength with time, from $3.8\pm0.5$\,\AA\ to $8.5\pm1.2$\,\AA; however, at 278\,d and 380\,d, this component is absent. The emission component at earlier times shows a Gaussian profile with an extra narrow emission line, caused by a contaminating \ion{H}{ii} region. At late times, the H$\alpha$ emission has a Lorentzian profile with a FWHM velocity of $815\pm65$\,km\,s$^{-1}$ at 295\,d, increasing to $980\pm55$\,km\,s$^{-1}$ at 403\,d. The absence of the absorption component, and the Lorentzian profile in emission, further indicate interaction between the ejecta and the CSM \citep{Chugai04b}. At 380\,d, on the top of the emission component of the H$\alpha$, a small notch is observed; upon close examination this was revealed to be residuals from the galaxy subtraction\footnote{The expansion velocities and the pseudo-equivalent-widths were measured removing the contribution of the host galaxy.}.   

Based on the width of the lines observed in the SN spectra, we infer very low expansion velocities. The velocity obtained for H$\alpha$ decreases from $\sim1500$\,km\,s$^{-1}$ at 80\,d, to $\sim1300$\,km\,s$^{-1}$ at 127\,d. The velocities found for other lines show a similar behavior: low expansion velocities ($<2000$\,km\,s$^{-1}$), and little evolution. 

\subsection{Comparison to other supernovae}
\label{sec:sncomp}

\begin{figure*}
\centering
\includegraphics[width=9cm]{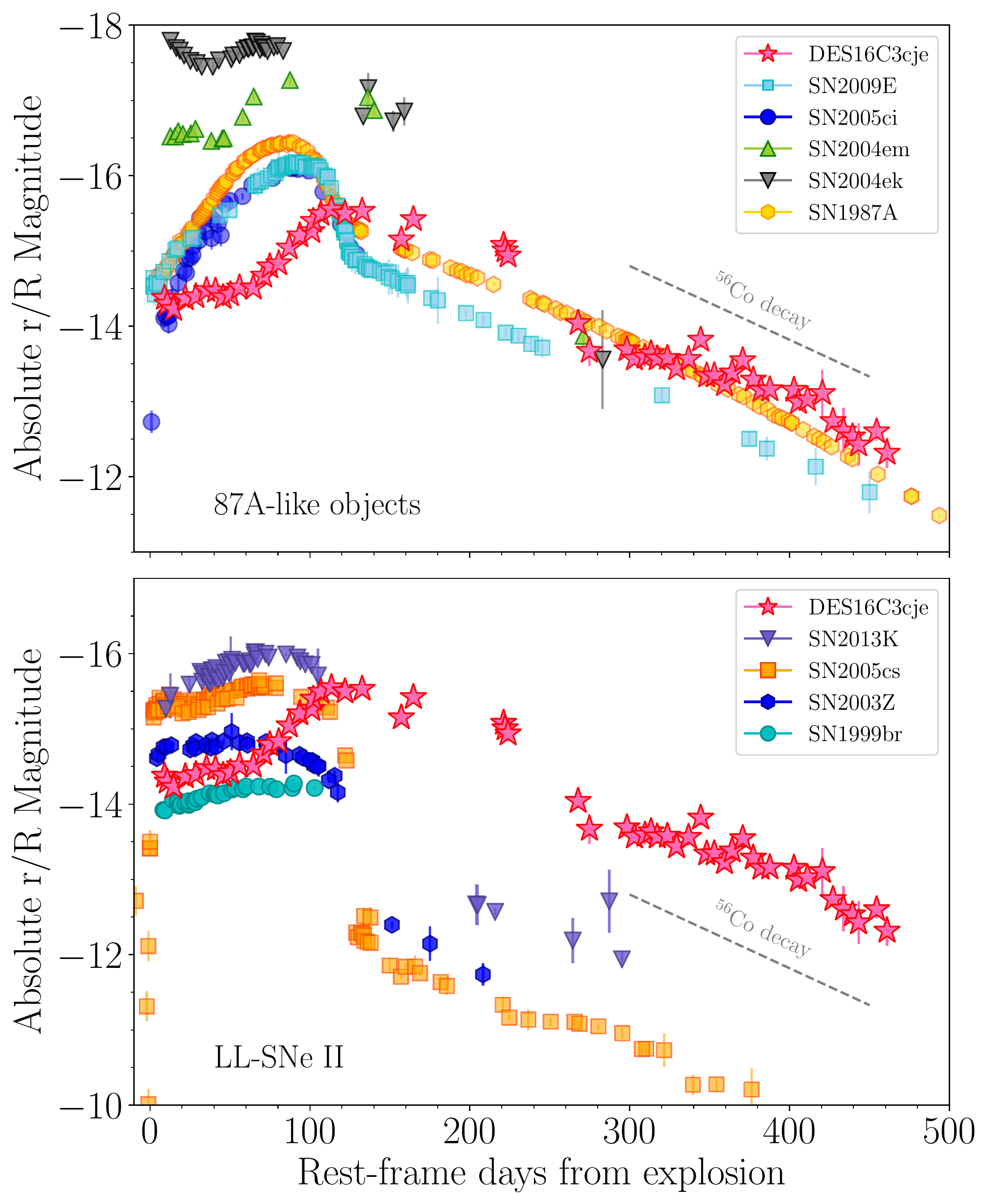}
\includegraphics[width=8.5cm]{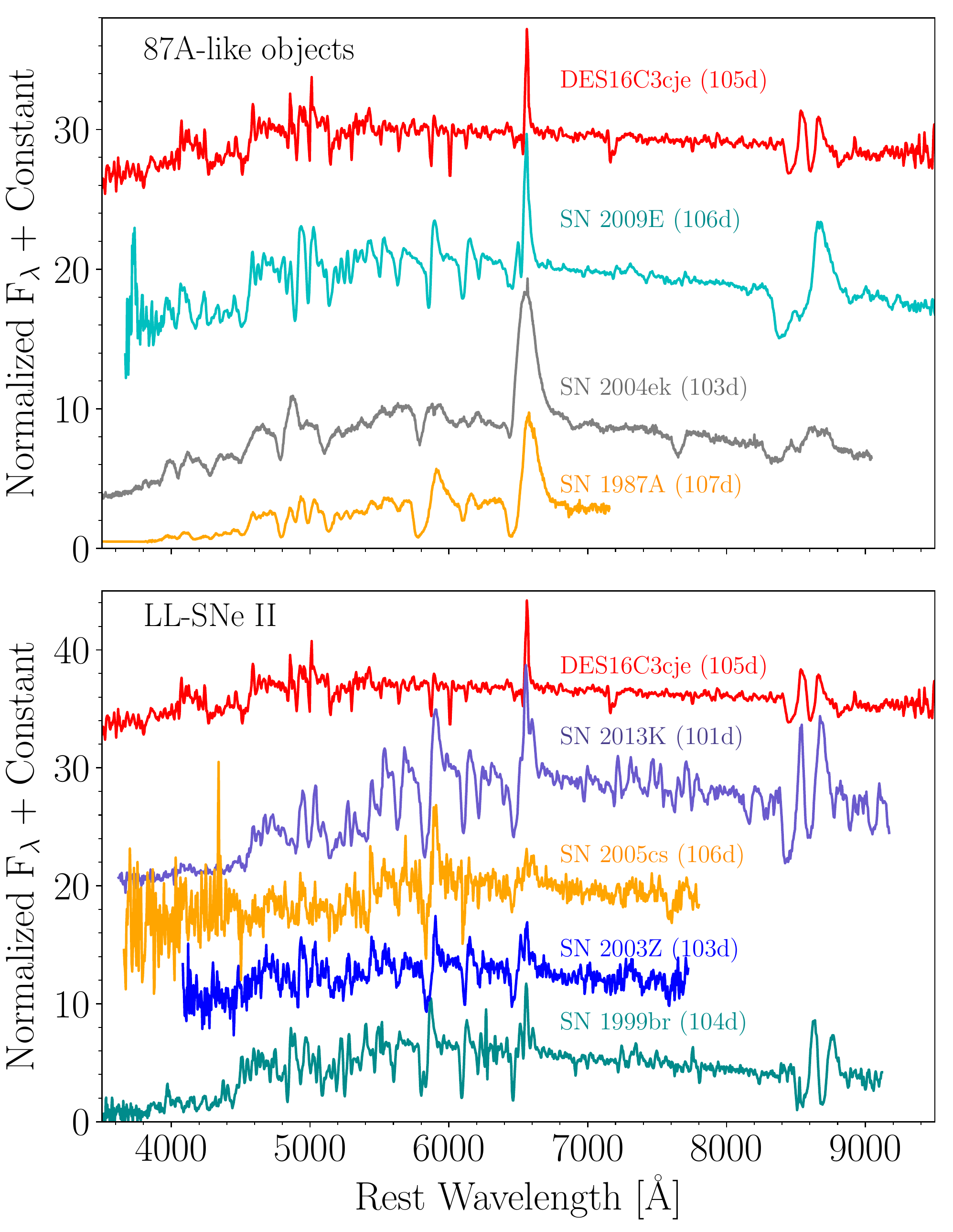}
\caption{Left: Comparison of the $r/R$-band light curve of DES16C3cje with well-observed SNe II. Upper panel: Comparison with 87A-like objects: SN~1987A \citep{Bouchet89,Hamuy90}, SN~2004ek \citep{Taddia16a}, SN~2004em \citep{Taddia16a}, SN~2005ci \citep{Taddia16a} and SN~2009E \citep{Pastorello12}. Lower panel: Comparison with LLSNe~II: SN~1999br \citep{Pastorello04, Galbany16}, SN2003Z \citep{Spiro14, Faran14a}, SN~2005cs \citep{Pastorello09}, and SN~2013K \citep{Tomasella18}. Right: Spectral comparison around 110\,d for DES16C3cje. Upper panel: Comparison with SN1987A-like objects: SN~1987A \citep{Phillips88}, SN~2004ek \citep{Taddia16a}, and SN~2009E \citep{Pastorello12}. Lower panel: Comparison with LLSNe~II: SN~1999br \citep{Gutierrez17a}, SN2003Z \citep{Spiro14}, SN~2005cs \citep{Pastorello09}, and SN~2013K \citep{Tomasella18}.}
\label{comp}
\end{figure*}

The slow rise of DES16C3cje is reminiscent of SN~1987A-like objects, whereas its low luminosity and low expansion velocities are a common characteristic in low luminosity (LL) SNe~II. In Figure~\ref{comp}, we show the photometric and spectral comparison of DES16C3cje with these two classes of events.  For the SN~1987A-like objects we compared with SN~1987A \citep{Bouchet89,Hamuy90}, which is the best observed and studied SN~II; SN~2004ek \citep{Taddia16a} and SN~2004em \citep{Taddia16a}, which both show a plateau before the main peak; SN~2005ci \citep{Taddia16a} and SN~2009E \citep{Pastorello12}, which are the faintest clones of SN~1987A. For the LL SNe~II, we select objects with spectra at around 110 days: SN~1999br \citep{Pastorello04,Galbany16,Gutierrez17a}, which is the faintest slowly-declining SN~II; SN2003Z \citep{Spiro14, Faran14a}, SN~2005cs \citep{Pastorello06,Pastorello09}, and SN~2013K \citep{Tomasella18}, which all have good photometric coverage in the first 150 days.
The long rise to peak is common between the SN1987A-like events and DES16C3cje; however, the rise is even longer for DES16C3cje. 

The full light curve evolution shows that DES16C3cje, from explosion to 60\,d, exhibits a initial \lq plateau\rq. Although this plateau is not common in SN1987A-like objects, two other SNe do show it: SN~2004ek (in the $V$ and $R$-bands) and SN~2004em (in the $I-$band, \citealt{Taddia16a}). \citet{Taddia16a} suggest that these two SNe are an intermediate case between SN~1987A and normal SNe~II. \citet{Pastorello12} argue that these plateaus are due to shock cooling. DES16C3cje also has the lowest luminosity within the SN1987A-like group, around 1\,mag fainter than SN~1987A and $\sim0.5$\,mag fainter than the low-luminosity SN~2009E.

Comparing to the LL-SNe~II sample, the initial evolution of DES16C3cje is consistent with typical SNe~II for $\sim60$\,d; however a sudden increase in luminosity transforms a \lq typical SN~II\rq\ to a SN1987A-like event. The post-peak light curve evolution also differs, where all SN1987A-like and LL-SNe follow the rate of \cofs\ decay. In the case of DES16C3cje, the decay at late-times is slower, again suggesting an extra source of energy is needed. We also note that SN~2005cs shows a slow decline soon after the plateau (between 140 and $\sim320$ days; \citealt{Pastorello09}). One possible explanation for this flattening was given by \citet{Utrobin07}, who suggested that it is produced by a residual contribution from radiation energy. Giving that this effect is predicted for typical slow-declining SNe II soon after the plateau phase, we explore an alternative scenario to explain the decay at the late-times in DES16C3cje. 

To distinguish between the scenarios of \cofs\ decay and accretion power (L $\propto t^{-5/3}$) as explanations for the light curves, we compare the reduced chi-squared ($\chi^2$) values (shown in Table~\ref{chi2}) of the corresponding fits to the SNe with data at late-time (between 280 and 500 days; DES16C3cje, SN~1987A, SN~2005cs and SN~2009E). Out of these, only for DES16C3cje does the power law provides a better fit ($\chi^2=0.71$), supporting the idea of an extra source of energy. Because of the large uncertainties in the bolometric light curve of DES16C3cje, we test this result using a Monte Carlo resampling with $10^5$ random draws (assuming a Gaussian distribution). The results obtained support our previous findings. 

Figure~\ref{comp} also presents the spectral comparison at $\sim105$\,d from explosion. The comparison with SN1987A-like objects and LL-SNe~II again shows that DES16C3cje is a unique object. None of the other SNe have lines as narrow as DES16C3cje. SN~1999br has the narrowest lines, but its spectrum also shows \ion{Ba}{ii} and \ion{Sc}{ii}, together with a multiple component H$\alpha$ P-Cygni profile, characteristic of LL-SNe~II.

\section{Light curve modelling}
\label{sec:modelling}

We now consider some models that can be used to understand and explain the physical origin and unusual features of DES16C3cje. For these models, we use the one-dimensional Lagrangian hydrodynamical code presented in \citet{Bersten11}. This code simulates a SN explosion, and produces bolometric light curves and photospheric velocities to characterize the progenitor and explosion properties. There are two particular challenges to this modelling: the early photometric behavior (before peak) and the low expansion velocities, and the late-time decline rate. We begin with the former.

There is a degree of degeneracy between the progenitor (pre-SN) mass and radius ($M$, $R$) and the explosion energy ($E$), which can be partially reduced by modeling the luminosity evolution together with the expansion velocity evolution. For DES16C3cje, the expansion velocities imply a low $E/M$ ratio. We found that for a progenitor with similar characteristics to those used for SN~1987A (i.e., a blue supergiant star with $R\sim50$\,\Rsun, $M_\mathrm{ZAMS}=20$\,\Msun\ and $E=1$\,foe), there is no model that simultaneously matches the light curve and velocity evolution, as a low energy is needed to reproduce the latter. The low energy required leads to a much fainter and broader light curve than that observed. We found that explosion energies of $\sim0.1$\,foe are needed to reproduce the expansion velocities of DES16C3cje.

Therefore, we calculated a grid of hydrodynamical models with values of $E$ close to $0.1$\,foe. Our pre-SN models were computed using the stellar evolution code MESA version 10398 \citep{Paxton11,Paxton13,Paxton15,Paxton18}. The stars were evolved from the pre-main-sequence to the time of core collapse, defined as when any part of the collapsing core exceeds an infall velocity of $1000$\,km\,s$^{-1}$, and assuming solar metallicity. Our models cover the $M_{\rm ZAMS}$ range of $9-25$~\Msun\ in intervals of 1~\Msun (which corresponds to progenitor radii  between 480 and 1050~\Rsun), and explosion energies between 0.1 and 0.5~foe with the exception of the largest masses and lower energies due to numerical difficulties.

\begin{figure*}
\centering
\includegraphics[width=\textwidth]{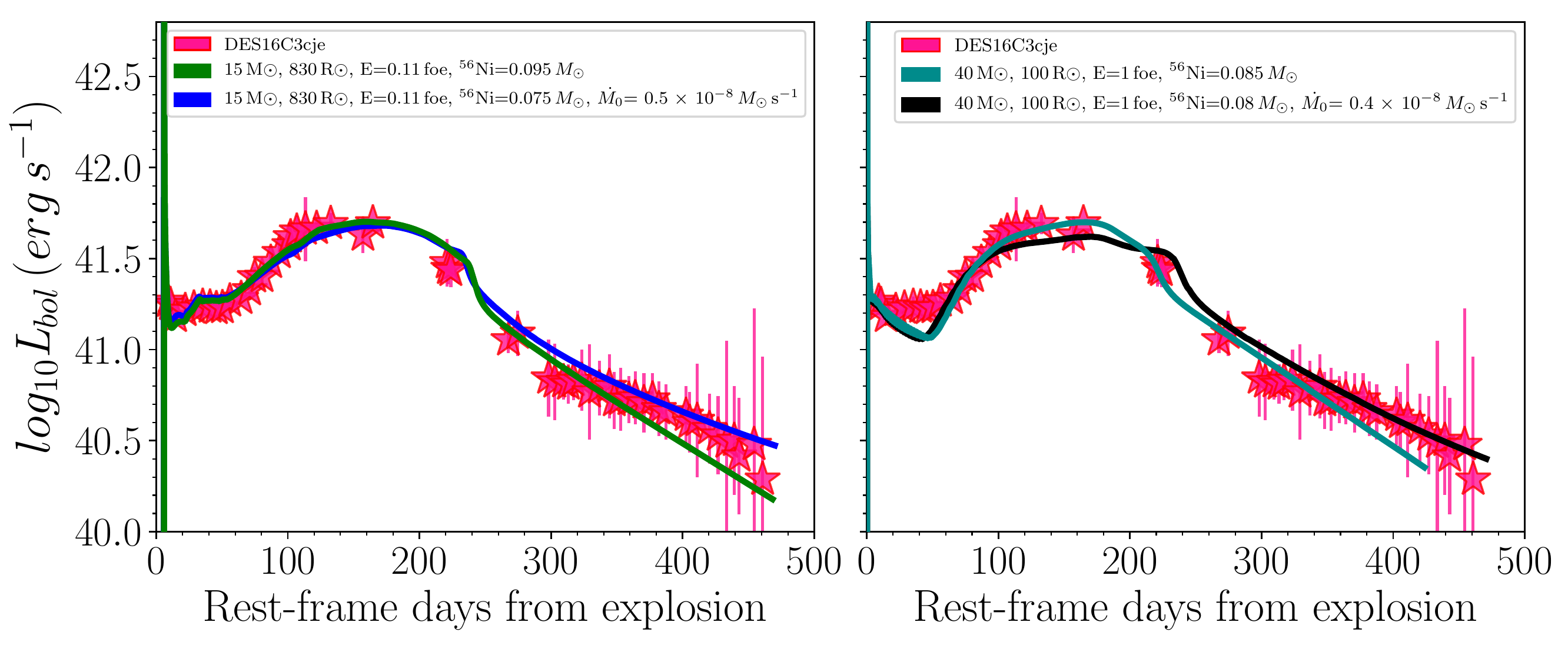}
\caption{Bolometric light curve of DES16C3cje (stars) compared with the results of the light curve calculations from hydrodynamic models (Section~\ref{sec:modelling}). A core-collapse explosion (powered only by \nifs) is presented in green on the left panel and in cyan on the right. Explosions powered by both \nifs\ and fallback (i.e., with an extra contribution of energy coming from accretion in the central region) are presented in blue and black in each panel. The parameters used for each model are given in the legends. 
}
\label{model}
\end{figure*}

After exploring several configurations (see Figure~\ref{fig:model15} in the Appendix), we found a model that reproduced the observations relatively well. This model is presented on the left panel of Figure~\ref{model} and has the following physical parameters: a $M_{\rm ZAMS}=15$\,\Msun, a pre-SN mass of $13.3$\,\Msun, $R=830$\,\Rsun\ and $E=0.11$\,foe. We also consider \nifs\ masses in the range of 0.01 and 0.1~\Msun\ and find that a relatively large \nifs\ mass of 0.095~\Msun\ is required to reproduce the light curve observed after the initial plateau. This material was mixed up to 0.75 of the pre-SN mass, and therefore a not too extreme mixing was required as is common in several 87A-like objects in order to produce the initial plateau and the long rise to the peak. In this scenario, the peculiar light curve shape of DES16C3cje can be understood as a combination of a low explosion energy and a relatively large \nifs\ production, while its progenitor has a red supergiant (RSG) structure typical of other SN II objects. 

We now turn to the late-time light curve. Despite the good agreement between the model and observations at early times, there are clear differences in the slopes during the light curve tail (green curve in Figure~\ref{model}). As discussed above, DES16C3cje does not follow the behavior expected by radioactive decay of \cofs, but instead is consistent with a power law $\propto t^{-5/3}$, compatible with the decline rate expected from accretion power (or \lq fallback\rq\ \footnote{The canonical power-law index, $n=-5/3$, is produced by a simple ballistic fallback model \citep{Rees88}. However, standard viscous disc descriptions extend the duration of the emission, with an index closer to $n=-1.2$ \citep[see][and references therein]{Balbus18}. This suggests that the index value $n$ changes depending on the conditions of the disc.}; \citealt{Michel88,Chevalier89}). Under some conditions, for example if the SN explosion is not powerful enough, some material may not acquire sufficient energy to escape and will eventually be accreted onto the compact remnant. These accretions are usually associated with powerful energy outflows. A fraction of this energy can be thermalised within the SN ejecta and thus power the light curve \citep{Dexter13}.

We have included this extra energy in our 1D Lagrangian code to explore if this can improve the differences between the model and observations during the latter part of the light curve. The rate input of energy due to the accretion can be written as: L$_{\rm fb}={\dot E}= \eta\, \dot{M}\, c^2$ where $\dot{M}$ is the fallback accretion rate, $c$ is the speed of light and $\eta$ is the efficiency factor, estimated to be of the order of $10^{-3}$ \citep{Dexter13}. Analytic estimates \citep{Chevalier89}, as well as numerical simulations \citep{Zhang08, Dexter13}, have shown that the accretion rate can be assumed to be $\dot{M}= \dot{M_0} (t/t_0)^{-5/3}$, where $\dot{M_0}$ is the accretion rate onto the remnant at a time $t_0$ when the fallback episode begins. The fallback energy is instantaneously deposited after the explosion, near the center of the progenitor, and we assume full trapping.

In our treatment, $\dot{M_0}$ and $t_0$ are free parameters to be determined by comparison with the observations.
We again calculate a grid of simulations, but this time vary $\dot{M}$ in the range of $10^{-7}-10^{-9}$ and $t_0$ between 0.1~d and 50~d after the onset of the simulation, finding a set of parameters that can reproduce the behaviour of the light-curve tail of DES16C3cje. In the lower panel of Figure~\ref{fig:model15}, we show the effect on the light curve and velocities as a result of the variation of $\dot{M_0}$, while in Figure~\ref{fig:modelt0}, the changes in the light curve produced by different $t_0$ are presented.
The fallback parameters found are: $\dot{M_0}=0.5\times 10^{-8}$\,$M_\odot$\,s$^{-1}$ and $t_0=1$\,d. These calculations were performed assuming the same progenitor and explosion energy as the RSG model presented above, and the combined model is shown in Figure~\ref{model} (left panel). The inclusion of fallback energy clearly improves the modelling during the tail, with almost no effect in other phases. However, we note a slightly smaller amount of \nifs\ is needed when fallback energy is added; a good match is found using 0.075\,$M_{\odot}$ of \nifs. The value of $\dot{M_0}$ is small compared with that usually found in the literature \citep{Zhang08,Moriya18}. The reason is the low luminosity of this SN: larger accretion rates inject more energy and produce brighter light curves. 

We emphasise that even though we try to model the light curve peak assuming that it was powered by fallback accretion instead of \nifs, we are unable to find any set of fallback parameters that can reproduce it. Larger accretion rates produce more luminous light curves and earlier plateaus than observed. In addition, a delayed deposition of the fallback energy is not a solution as despite the low accretion rate, a time delay factor produces an extremely luminous plateau (similar to figure~2 of \citealt{Moriya19}) and a brighter light curve tail.

The parameters of our preferred model point to a normal RSG progenitor that has experienced a low energy explosion leading to the fallback process. The peculiar light curve shape of DES16C3cje can then be explained as a combination of a low explosion energy, a relatively large \nifs\ mass but not extremely mixing, and extra energy due to the accretion of material onto the compact remnant.

There is strong evidence of the existence of a correlation between the explosion energy and the amount of $^{56}$Ni \citep[see for example][]{Pejcha15}, in the sense that more energetic events produce larger amount of $^{56}$Ni. This relation is also supported by theoretical studies. The low explosion energy and the relatively large \nifs\ production found in our modelling does not follow the expected correlation. We note a low explosion energy was mainly required to reproduce the \textnormal{low-expansion} velocities.

DES16C3cje has only two measurements of the expansion velocity available at $\sim105$\,d and $\sim127$\,d, and thus the expansion velocity during the first weeks of evolution is not unambiguously known, and the measurements around $\sim100$\,d may not represent the photospheric velocities of the ejecta. We experiment with relaxing the condition to reproduce the expansion velocity, and find an alternative model that reproduces relatively well the observed light curve with a progenitor with $\sim$40\,\Msun, an explosion energy of 1\,foe and $0.08$\,\Msun\ of $^{56}$Ni (Figure~\ref{model}). Here, we used a polytropic model to describe the structure of the star before explosion. The fallback parameters needed to reproduce the tail are similar to that in the previous model, i.e, $\dot{M_0}=0.4\times 10^{-8}$\,\Msun\,s$^{-1}$ and $t_0=1$\,d. The higher energy of this model is then more consistent with known correlations between \nifs\ production and explosion energy.
Figure~\ref{fig:model40} shows the different configurations explored for this case. 
The parameters of the best-fit models are presented in Table~\ref{params}.

\section{Discussion and conclusions}
\label{sec:summary}

DES16C3cje is a low-luminosity and low-velocity type II supernova (SN~II). Its light curves show a plateau for $\sim60$ days, followed by a long rise time, reminiscent of SN~1987A, but on a longer time-scale. The initial faint plateau can be explained by hydrogen recombination, while the broad peak is powered by radioactive decay. After 300 days, the tail presents a decline rate comparable to that expected from accretion power ($\propto t^{-5/3}$). The narrow lines observed in the spectra imply low expansion velocities, and thus, low explosion energies. Taken together, these characteristics suggest an unusual explosion.   

Modelling the light curve of DES16C3cje and its velocity evolution with hydrodynamical calculations, we have shown that the SN is consistent with the explosion of a RSG star with a mass of 15\,\Msun, an energy of $0.11$\,foe, and synthesising a \nifs\ mass of 0.075\,\Msun. Because of the low energy in the explosion, some material is accreted by the compact remnant with an accretion rate of $\sim0.5\times{10^{-8}}$ M$\odot$ s$^{-1}$.  Although this scenario reproduces the light curve and velocities, at first sight the required \nifs\ mass appears relatively large for two main reasons: 1) low energy explosions are observed to produce small amounts of \nifs, and 2) in the fallback scenario, some amount of the \nifs\ is expected to be accreted on to the central remnant. 

However, \citet{Chevalier89} discussed the expectation that an ejection of substantial $^{56}$Ni would imply little mass fallback, and showed this is not valid for accretion after the passage of the reverse shock wave, when the $^{56}$Ni is expected to mix with outer core layers. \citet{Heger10} further showed that a considerable amount of $^{56}$Ni comes out when mixing precedes fallback. The mixing in RSGs is larger than in compact objects as perturbations have more time to grow before freezing out. Under these considerations, it is not unusual to find SNe that both experienced some fallback and have a relatively large amount of $^{56}$Ni. 

Nonetheless, we also consider an alternative scenario by assuming that the velocities measured from the absorption lines at 105 and 126 days do not represent the photospheric velocities of the ejecta. We then find that DES16C3cje can be modelled as the explosion of a relatively compact star ($R=100$\,\Rsun), with a mass of $\sim40$\,\Msun, an explosion energy of $1$\,foe, and a \nifs\ mass of $0.08$\,\Msun. 

Both models can reproduce the overall evolution of the light curve of DES16C3cje; however, the low-energy explosion of a RSG fits the early part of the light curve better, and provides a good agreement with expansion velocities.

A further possibility to explain the late-time light curve of DES16C3cje is interaction with CSM. Interacting objects \citep[e.g. SNe~IIn, SN 2009ip-like objects;][]{Stritzinger12, Fraser15, Elias-Rosa16,  Pastorello18} often have flattened late-time light curves, with decline rates slower than that expected for $^{56}$Co decay. The flat evolution in the light curves of DES16C3cje, together with the lack of \ion{[O}{ii]} $\lambda\lambda$6300, 6363, \ion{[Fe}{ii]} $\lambda7155$ and \ion{[Ca}{ii]} $\lambda$7291, 7323 emission lines, offer some support for this scenario. However, this evidence for interaction only appears at around 300 days from explosion with no evidence for interaction prior to this epoch, in turn suggesting a significant mass loss during the progenitor star evolution. 

Theoretical models have also shown that stars with masses below 40 \Msun\ at low-metallicities undergo very little mass loss due to stellar winds \citep[e.g.][]{Woosley07,Meynet13}. Assuming that the progenitor mass favored by our hydrodynamical models (15 and 40\Msun) is correct, we would expect a low  mass loss. The location of our object supports this argument: DES16C3cje exploded in a low-luminosity (low-metallicity, Sec.~\ref{sec:host}) host, and models predict low-metallicity stars have less mass loss and bigger hydrogen envelopes when they explode \citep[e.g.][]{Heger03}.

While the late-time light curve of DES16C3cje is following a decline rate close to $t^{-5/3}$, we cannot rule out a scenario involving interaction with CSM. \citet{Moriya19} briefly discuss the possibility of CSM interaction in fallback SN and the need to study this issue in the future.

In summary, we have shown that the fallback SN scenario can naturally explain the slow decline in the late-time light curve. However, further investigations are needed to interpret the origin of these peculiar objects, the signatures required to identify the explosion scenario, and the role of the \nifs\ mass and interaction with CSM.

\section*{Acknowledgements}
We are grateful to Pedro Lacerda for performing some of the observations used in this work.
We thank the anonymous referee for useful suggestions.

CPG and MS acknowledge support from EU/FP7-ERC grant No. [615929]. LG was funded by the European Union's Horizon 2020 research and innovation programme under the Marie Sk\l{}odowska-Curie grant agreement No. 839090.
TWC acknowledgments the funding provided by the Alexander von Humboldt Foundation. 
MF is supported by a Royal Society -- Science Foundation Ireland University Research Fellowship.
MG is supported by the Polish NCN MAESTRO grant 2014/14/A/ST9/00121. 
MN is supported by a Royal Astronomical Society Research Fellowship.

This work used data collected at the European Organisation for Astronomical Research in the Southern Hemisphere, Chile, under program IDs: 299.D-5040(A), 299.D-5040(B), 0100.D-0461(A), 194.C-0207(I), and as part of PESSTO, (the Public ESO Spectroscopic Survey for Transient Objects Survey) ESO program 197.D-1075, 199.D-0143.

Some of the data presented here were obtained at the Gemini Observatory, which is operated by the Association of Universities for Research in Astronomy, Inc., under a cooperative agreement with the NSF on behalf of the Gemini partnership: the National Science Foundation (United States), the National Research Council (Canada), CONICYT (Chile), Ministerio de Ciencia, Tecnolog\'ia e Innovaci\'on Productiva (Argentina), and Minist\'erio da Ci\^encia, Tecnologia e Inova\c{c}\~ao (Brazil). Gemini observations were obtained under programme NOAO GS-2016B-Q-9.

Funding for the DES Projects has been provided by the U.S. Department of Energy, the U.S. National Science Foundation, the Ministry of Science and Education of Spain, the Science and Technology Facilities Council of the United Kingdom, the Higher Education Funding Council for England, the National Center for Supercomputing Applications at the University of Illinois at Urbana-Champaign, the Kavli Institute of Cosmological Physics at the University of Chicago, the Center for Cosmology and Astro-Particle Physics at the Ohio State University, the Mitchell Institute for Fundamental Physics and Astronomy at Texas A\&M University, Financiadora de Estudos e Projetos, Funda{\c c}{\~a}o Carlos Chagas Filho de Amparo {\`a} Pesquisa do Estado do Rio de Janeiro, Conselho Nacional de Desenvolvimento Cient{\'i}fico e Tecnol{\'o}gico and the Minist{\'e}rio da Ci{\^e}ncia, Tecnologia e Inova{\c c}{\~a}o, the Deutsche Forschungsgemeinschaft and the Collaborating Institutions in the Dark Energy Survey. 

The Collaborating Institutions are Argonne National Laboratory, the University of California at Santa Cruz, the University of Cambridge, Centro de Investigaciones Energ{\'e}ticas, Medioambientales y Tecnol{\'o}gicas-Madrid, the University of Chicago, University College London, the DES-Brazil Consortium, the University of Edinburgh, the Eidgen{\"o}ssische Technische Hochschule (ETH) Z{\"u}rich, Fermi National Accelerator Laboratory, the University of Illinois at Urbana-Champaign, the Institut de Ci{\`e}ncies de l'Espai (IEEC/CSIC), the Institut de F{\'i}sica d'Altes Energies, Lawrence Berkeley National Laboratory, the Ludwig-Maximilians Universit{\"a}t M{\"u}nchen and the associated Excellence Cluster Universe, the University of Michigan, the National Optical Astronomy Observatory, the University of Nottingham, The Ohio State University, the University of Pennsylvania, the University of Portsmouth, SLAC National Accelerator Laboratory, Stanford University, the University of Sussex, Texas A\&M University, and the OzDES Membership Consortium.

Based in part on observations at Cerro Tololo Inter-American Observatory, National Optical Astronomy Observatory, which is operated by the Association of Universities for Research in Astronomy (AURA) under a cooperative agreement with the National Science Foundation.

Part of the funding for GROND (both hardware as well as personnel) was generously granted from the Leibniz-Prize to Prof. G. Hasinger (DFG grant HA 1850/28-1).

This paper includes data gathered with the 6.5 meter Magellan Clay Telescope located at Las Campanas Observatory, Chile under the programme 100type  IA  Supernovae (100IAS) survey. 

This work has been partially supported by the Spanish grant PGC2018-095317-B-C21 within the European Funds for Regional Development (FEDER).

The DES data management system is supported by the National Science Foundation under Grant Numbers AST-1138766 and AST-1536171. The DES participants from Spanish institutions are partially supported by MINECO under grants AYA2015-71825, ESP2015-66861, FPA2015-68048, SEV-2016-0588, SEV-2016-0597, and MDM-2015-0509, some of which include ERDF funds from the European Union. IFAE is partially funded by the CERCA program of the Generalitat de Catalunya. Research leading to these results has received funding from the European Research Council under the European Union's Seventh Framework Program (FP7/2007-2013) including ERC grant agreements 240672, 291329, and 306478. We  acknowledge support from the Brazilian Instituto Nacional de Ci\^enciae Tecnologia (INCT) e-Universe (CNPq grant 465376/2014-2).

This research used resources of the National Energy Research Scientific Computing Center (NERSC), a U.S. Department of Energy Office of Science User Facility operated under Contract No. DE-AC02-05CH11231.

This manuscript has been authored by Fermi Research Alliance, LLC under Contract No. DE-AC02-07CH11359 with the U.S. Department of Energy, Office of Science, Office of High Energy Physics. The United States Government retains and the publisher, by accepting the article for publication, acknowledges that the United States Government retains a non-exclusive, paid-up, irrevocable, world-wide license to publish or reproduce the published form of this manuscript, or allow others to do so, for United States Government purposes.




\bibliography{Bibliography}
\bibliographystyle{mnras}



\appendix

\section{Tables}

\renewcommand{\thetable}{A\arabic{table}}
\setcounter{table}{0}
\begin{table*}
\centering
\small
\caption{Photometry of DES16C3cje}
\label{photo}
\begin{tabular}[t]{ccccccccl}
\hline
UT date    &	MJD	&  Rest-frame phase   &       $g$	     &	$r$	       &	$i$	 &       $z$       &   Instrument     \\
           &            &  (days)  &     (mag)       &     (mag)       &     (mag)       &     (mag)       &                  \\
\hline
\hline               
20161011   &  57673.3	&   2.9    &  \nodata        &  \nodata        &  \nodata        &   $23.26\pm0.08$ &  DECam      \\
20161018   &  57680.3	&   9.5    &  \nodata        &  $22.71\pm0.13$ &  \nodata        &   \nodata        &  DECam      \\
20161019   &  57681.1	&   10.3   &  $23.31\pm0.24$ &  \nodata        &  $22.86\pm0.26$ &   $23.01\pm0.08$ &  DECam      \\
20161020   &  57682.1	&   11.2   &  \nodata        &  $22.78\pm0.06$ &  $22.87\pm0.05$ &   \nodata        &  DECam      \\
20161024   &  57686.3	&   15.2   &  $23.22\pm0.08$ &  $22.85\pm0.05$ &  \nodata        &   $22.86\pm0.08$ &  DECam      \\
20161025   &  57687.2	&   16.0   &  $23.24\pm0.08$ &  \nodata        &  $22.91\pm0.07$ &   \nodata        &  DECam      \\
20161101   &  57694.1	&   22.5   &  $23.23\pm0.07$ &  $22.71\pm0.03$ &  $22.71\pm0.06$ &   $22.73\pm0.06$ &  DECam      \\
20161108   &  57701.1	&   29.1   &  $23.21\pm0.08$ &  $22.68\pm0.03$ &  $22.67\pm0.03$ &   $22.62\pm0.03$ &  DECam      \\
20161115   &  57708.2	&   35.8   &  $23.28\pm0.10$ &  $22.60\pm0.04$ &  $22.61\pm0.05$ &   $22.54\pm0.05$ &  DECam      \\
20161121   &  57714.1	&   41.3   &  $23.43\pm0.13$ &  $22.61\pm0.05$ &  $22.57\pm0.04$ &   $22.50\pm0.04$ &  DECam      \\
20161122   &  57715.1	&   42.3   &  $23.84\pm0.18$ &  \nodata        &  \nodata        &   \nodata        &  DECam      \\
20161123   &  57716.2	&   43.3   &  $23.42\pm0.09$ &  \nodata        &  \nodata        &   \nodata        &  DECam      \\
20161126   &  57719.2	&   46.1   &  $23.42\pm0.08$ &  $22.70\pm0.03$ &  $22.54\pm0.03$ &   $22.48\pm0.03$ &  DECam      \\
20161201   &  57724.1	&   50.8   &  $23.47\pm0.08$ &  $22.66\pm0.03$ &  $22.54\pm0.05$ &   $22.46\pm0.04$ &  DECam      \\
20161207   &  57730.1	&   56.4   &  $23.42\pm0.17$ &  $22.57\pm0.03$ &  $22.50\pm0.04$ &   $22.31\pm0.04$ &  DECam      \\
20161216   &  57739.2	&   65.0   &  $23.37\pm0.15$ &  $22.57\pm0.03$ &  $22.40\pm0.03$ &   $22.28\pm0.04$ &  DECam      \\
20161223   &  57746.1	&   71.5   &  $23.26\pm0.04$ &  $22.41\pm0.02$ &  $22.27\pm0.02$ &   $22.24\pm0.02$ &  DECam      \\
20161227   &  57750.2	&   75.3   &  $23.12\pm0.06$ &  $22.29\pm0.03$ &  $22.10\pm0.04$ &   \nodata        &  DECam      \\
20161228   &  57751.1	&   76.2   &  \nodata        &  \nodata        &  $22.25\pm0.03$ &   $22.11\pm0.03$ &  DECam      \\
20170102   &  57756.1	&   80.9   &  $23.02\pm0.06$ &  $22.25\pm0.02$ &  $22.11\pm0.02$ &   $22.02\pm0.03$ &  DECam      \\
20170109   &  57763.1	&   87.5   &  \nodata        &  $22.04\pm0.03$ &  $21.96\pm0.02$ &   \nodata        &  DECam      \\
20170116   &  57770.1	&   94.1   &  $22.60\pm0.03$ &  $21.88\pm0.01$ &  $21.85\pm0.01$ &   $21.76\pm0.02$ &  DECam      \\
20170121   &  57775.2	&   98.9   &  $22.51\pm0.04$ &  \nodata        &  \nodata        &   \nodata        &  DECam      \\
20170124   &  57778.1	&   101.6  &  \nodata        &  $21.82\pm0.02$ &  \nodata        &   $21.65\pm0.02$ &  DECam      \\
20170125   &  57779.1	&   102.6  &  $22.31\pm0.04$ &  \nodata        &  $21.68\pm0.01$ &   \nodata        &  DECam      \\
20170128   &  57782.1	&   105.4  &  \nodata        &  $21.74\pm0.02$ &  \nodata        &   $21.58\pm0.02$ &  DECam      \\
20170130   &  57784.1	&   107.3  &  $22.27\pm0.03$ &  \nodata        &  $21.59\pm0.02$ &   \nodata        &  DECam      \\
20170204   &  57789.1	&   112.0  &  \nodata        &  $21.73\pm0.05$ &  \nodata        &   $21.54\pm0.02$ &  DECam      \\
20170206   &  57791.1	&   113.9  &  $22.62\pm0.19$ &  \nodata        &  $21.53\pm0.02$ &   \nodata        &  DECam      \\
20170215   &  57800.1	&   122.3  &  $22.09\pm0.03$ &  $21.58\pm0.02$ &  \nodata        &   \nodata        &  DECam      \\
20170218   &  57803.5	&   125.5  &  \nodata        &  \nodata        &  $21.39\pm0.01$ &   \nodata        &  DECam      \\
20170227   &  57811.5	&   133.1  &  \nodata        &  $21.55\pm0.20$ &  \nodata        &   \nodata        &  EFOSC2     \\
20170325   &  57837.5	&   157.6  &  $21.94\pm0.10$ &  $21.93\pm0.10$ &  $20.90\pm0.10$ &   \nodata        &  EFOSC2     \\  
20170402   &  57845.5	&   165.1  &  \nodata        &  $21.66\pm0.22$ &  \nodata        &   \nodata        &  EFOSC2     \\  
20170412   &  57855.5	&   174.5  &  $>20.47$       &  $>20.85$       &  $>20.52$       &   $>20.68$       &  GROND      \\  
20170531   &  57905.4	&   221.5  &  \nodata        &  $22.00\pm0.10$ &  \nodata        &   \nodata        &  EFOSC2     \\
20170601   &  57906.4	&   222.5  &  \nodata        &  $22.07\pm0.22$ &  \nodata        &   \nodata        &  EFOSC2     \\
20170603   &  57908.4	&   224.3  &  \nodata        &  $22.14\pm0.15$ &  $21.55\pm0.15$ &   \nodata        &  EFOSC2     \\
20170720   &  57955.9	&   269.1  &  \nodata        &  $23.04\pm0.20$ &  $22.88\pm0.06$ &   \nodata        &  LDSS3      \\
20170727   &  57962.4	&   275.2  &  $22.93\pm0.05$ &  $23.41\pm0.30$ &  $23.27\pm0.31$ &   $23.30\pm0.30$ &  FORS2      \\  
20170821   &  57987.3	&   298.6  &  $24.87\pm0.27$ &  $23.39\pm0.06$ &  $23.76\pm0.10$ &   $23.36\pm0.09$ &  DECam      \\
20170826   &  57992.3	&   303.4  &  $24.80\pm0.23$ &  $23.51\pm0.06$ &  $23.81\pm0.11$ &   $23.33\pm0.12$ &  DECam      \\
\hline                
\end{tabular}
\end{table*}

\begin{table*}
\centering
\small
\setcounter{table}{0}
\begin{tabular}[t]{ccccccccl}
\hline
UT date    &	MJD	&  Rest-frame phase   &       $g$	     &	$r$	       &	$i$	 &       $z$       &  Instrument  \\
           &            &  (days)  &     (mag)       &     (mag)       &     (mag)       &     (mag)       &              \\
\hline
\hline
20170831   &  57997.3	&   308.1  &  \nodata        &  \nodata        &  $23.76\pm0.31$ &   \nodata        &  DECam      \\
20170901   &  57998.2	&   308.9  &  \nodata        &  \nodata        &  $23.53\pm0.24$ &   $23.11\pm0.24$ &  DECam      \\
20170902   &  57999.3	&   309.9  &  \nodata        &  $23.49\pm0.13$ &  \nodata        &   $23.47\pm0.19$ &  DECam      \\
20170906   &  58003.3	&   313.7  &  \nodata        &  $23.44\pm0.17$ &  $23.95\pm0.19$ &   $23.37\pm0.08$ &  DECam      \\
20170910   &  58007.4	&   317.6  &  \nodata        &  $23.51\pm0.09$ &  $23.63\pm0.12$ &   \nodata        &  DECam      \\
20170912   &  58009.4	&   319.5  &  \nodata        &  \nodata        &  $23.68\pm0.12$ &   $23.35\pm0.06$ &  DECam      \\
20170917   &  58014.2	&   324.0  &  $24.68\pm0.20$ &  $23.50\pm0.05$ &  \nodata        &   \nodata        &  DECam      \\
20170923   &  58020.3	&   329.7  &  $24.88\pm0.28$ &  $23.64\pm0.08$ &  $24.06\pm0.10$ &   $23.33\pm0.06$ &  DECam      \\
20171001   &  58028.2	&   337.2  &  \nodata        &  $23.52\pm0.11$ &  \nodata        &   $23.37\pm0.09$ &  DECam      \\
20171009   &  58036.3	&   344.8  &  \nodata        &  $23.26\pm0.11$ &  $23.93\pm0.12$ &   $23.48\pm0.09$ &  DECam      \\
20171013   &  58040.3	&   348.6  &  \nodata        &  $23.74\pm0.07$ &  \nodata        &   $23.66\pm0.10$ &  DECam      \\
20171018   &  58045.3	&   353.3  &  $24.90\pm0.25$ &  $23.74\pm0.07$ &  $24.04\pm0.12$ &   $23.64\pm0.11$ &  DECam      \\
20171025   &  58052.1	&   359.7  &  $25.00\pm0.24$ &  $23.85\pm0.10$ &  $24.00\pm0.15$ &   $23.54\pm0.10$ &  DECam      \\
20171030   &  58057.2	&   364.5  &  \nodata        &  $23.70\pm0.09$ &  $24.02\pm0.12$ &   $23.55\pm0.08$ &  DECam      \\
20171106   &  58064.2	&   371.1  &  \nodata        &  $23.53\pm0.12$ &  $24.19\pm0.16$ &   $23.67\pm0.11$ &  DECam      \\
20171113   &  58071.2	&   377.7  &  \nodata        &  $23.79\pm0.06$ &  $23.93\pm0.10$ &   $23.62\pm0.09$ &  DECam      \\
20171118   &  58076.3	&   382.5  &  $25.17\pm0.32$ &  $23.93\pm0.10$ &  $24.19\pm0.18$ &   \nodata        &  DECam      \\
20171121   &  58079.1	&   385.1  &  \nodata        &  \nodata        &  \nodata        &   $23.79\pm0.12$ &  DECam      \\
20171124   &  58082.2	&   388.0  &  \nodata        &  $23.92\pm0.08$ &  $24.34\pm0.17$ &   \nodata        &  DECam      \\
20171126   &  58084.3	&   390.0  &  $25.02\pm0.36$ &  \nodata        &  \nodata        &   $23.75\pm0.11$ &  DECam      \\
20171204   &  58092.2	&   397.4  &  \nodata        &  \nodata        &  \nodata        &   $23.90\pm0.15$ &  DECam   \\
20171210   &  58098.2	&   403.1  &  $24.56\pm0.24$ &  $23.93\pm0.10$ &  $24.33\pm0.17$ &   $23.85\pm0.14$ &  DECam   \\
20171213   &  58101.2	&   405.9  &  \nodata        &  $24.09\pm0.10$ &  $24.49\pm0.17$ &   \nodata        &  DECam   \\
20171219   &  58107.2	&   411.6  &  $24.97\pm0.24$ &  $24.05\pm0.07$ &  $24.73\pm0.16$ &   $23.96\pm0.13$ &  DECam   \\
20171229   &  58117.1	&   420.9  &  \nodata        &  $23.97\pm0.31$ &  $24.70\pm0.32$ &   $24.01\pm0.22$ &  DECam   \\
20180105   &  58124.1	&   427.5  &  \nodata        &  $24.34\pm0.08$ &  $24.65\pm0.14$ &   $24.05\pm0.11$ &  DECam   \\
20180112   &  58131.1	&   434.1  &  \nodata        &  $24.47\pm0.30$ &  $25.17\pm0.33$ &   $24.11\pm0.12$ &  DECam   \\
20180118   &  58137.1	&   439.7  &  \nodata        &  $24.55\pm0.14$ &  \nodata        &   $24.37\pm0.24$ &  DECam   \\
20180122   &  58141.1	&   443.5  &  \nodata        &  $24.65\pm0.28$ &  \nodata        &   \nodata        &  DECam   \\
20180203   &  58153.1   &   454.8  &  \nodata        &  $24.48\pm0.12$ &  \nodata        &   \nodata        &  DECam   \\ 
20180210   &  58160.0   &   461.3  &  \nodata        &  $24.76\pm0.20$ &  \nodata        &   \nodata        &  DECam   \\ 
\hline                
\hline
\hline
\end{tabular}
\begin{list}{}{}
\item \textbf{Notes:} The magnitudes have not been corrected for extinction.
\textbf{DECam:} Dark Energy Camera at Blanco 4-m telescope;
\textbf{EFOSC2:} ESO Faint Object Spectrograph and Camera at the 3.5-m ESO New Technology Telescope (NTT);
\textbf{GROND:} Gamma-Ray Burst Optical/Near-Infrared Detector at the 2.2-m MPG telescope;
\textbf{LDSS3:} Low Dispersion Survey Spectrograph at the Magellan Clay 6.5-m telescope;
\textbf{FORS2}: FOcal Reducer/low dispersion Spectrograph 2 at the ESO Very Large Telescope (VLT).
\end{list}
\end{table*}


\renewcommand{\thetable}{A\arabic{table}}
\setcounter{table}{1}
\begin{table*}
\centering
\small
\caption{$\chi^2$ for the power-law and exponential fits at late-time (between 280 and 500 days from explosion).}
\label{chi2}
\begin{tabular}[t]{ccccl}
\hline
SN                  &  $\chi^2$ Power-law   & $\chi^2$ Exponential  \\
                    &  (accretion power)    &  (\cofs\ decay)       \\ 
\hline
\hline 
DES16C3cje  &       0.710           &   2.384               \\
SN~1987A    &       14.060          &   2.551               \\
SN~2005cs   &       5.871           &   0.249               \\
SN~2009E    &       2.510           &   0.116              \\
\hline
\hline     
\end{tabular}
\end{table*}

\renewcommand{\thetable}{A\arabic{table}}
\setcounter{table}{2}
\begin{table*}
\centering
\small
\caption{Parameters of the best models presented in Figure~\ref{model}.}
\label{params}
\begin{tabular}[t]{ccccccccl}
\hline
Model &   Mass  & Radius  &  Energy  &  Ni mass  &   $\dot{M_0}$       & Reference  \\
      & (\Msun) & (\Rsun) &  (Foe)   &  (\Msun)  & (\Msun\,s$^{-1}$)   & (Colour) \\ 
\hline
\hline 
RSG   &  15     & 830     &  0.11    &  0.095    &   ...               & Green line \\
RSG   &  15     & 830     &  0.11    &  0.095    & $0.5\times 10^{-8}$ & Blue line  \\
\hline
BSG   &  40     & 100     &  1.0     &  0.085    &   ...               & Cyan line   \\
BSG   &  40     & 100     &  1.0     &  0.080    & $0.4\times 10^{-8}$ & Black line  \\       \hline
\hline     
\end{tabular}
\end{table*}

\clearpage

\section{Figures}

\setcounter{figure}{0}
\renewcommand\thefigure{\thesection.\arabic{figure}}  
\begin{figure*}
\centering
\includegraphics[width=\textwidth]{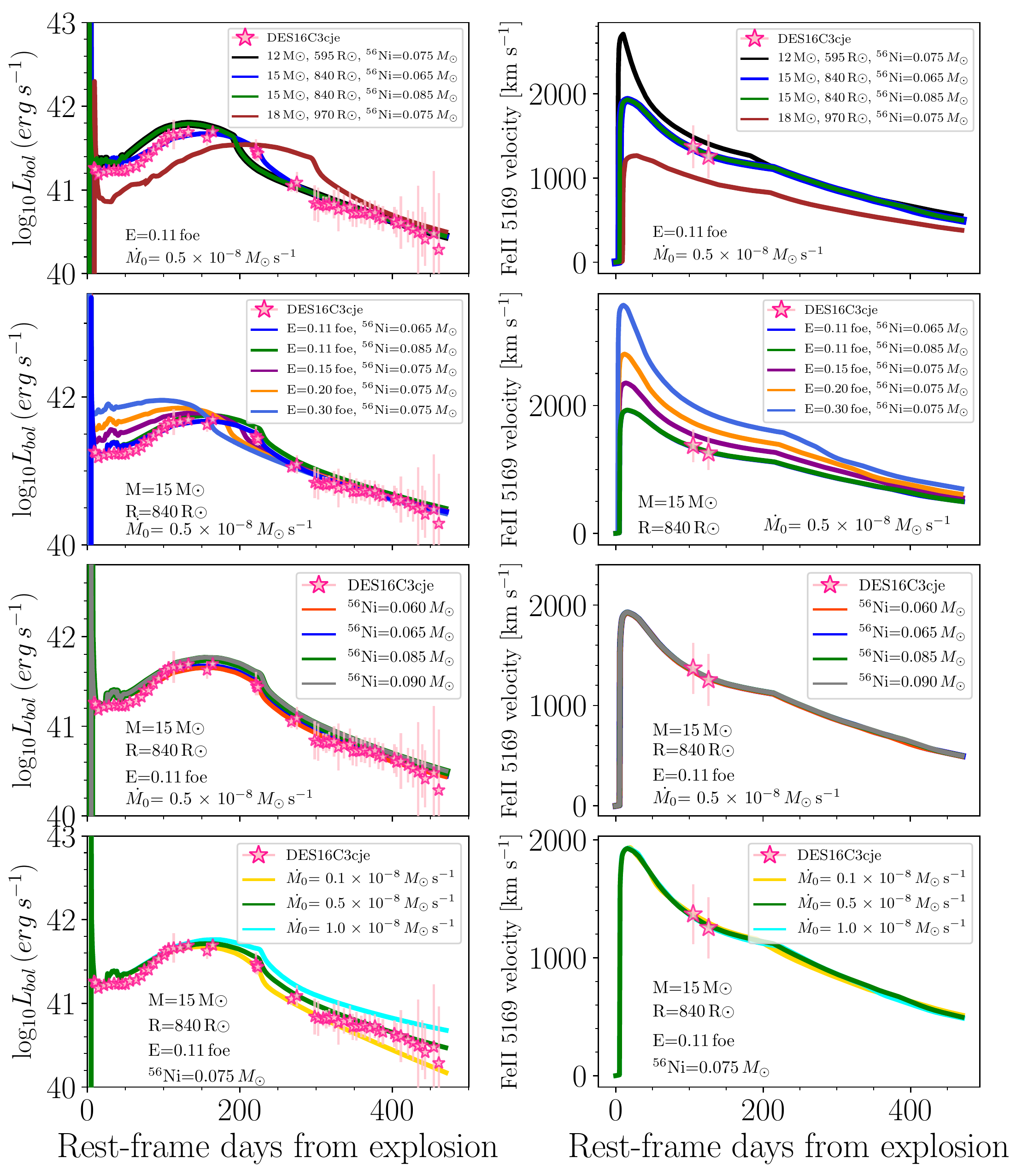}
\caption{\textbf{Left:} Bolometric light curve of DES16C3cje (stars) compared with the results of the light curve calculations from hydrodynamic models. For each plot, the legend shows the differences in the models, while the parameters with similar values are presented next to the curves. \textbf{Right}: 
Evolution of the photospheric velocity for the models presented in the left panel compared with measured Fe II 5169 \AA line velocities of DES16C3cje.}
\label{fig:model15}
\end{figure*}

\setcounter{figure}{1}
\renewcommand\thefigure{\thesection.\arabic{figure}}  
\begin{figure*}
\centering
\includegraphics[width=0.5\textwidth]{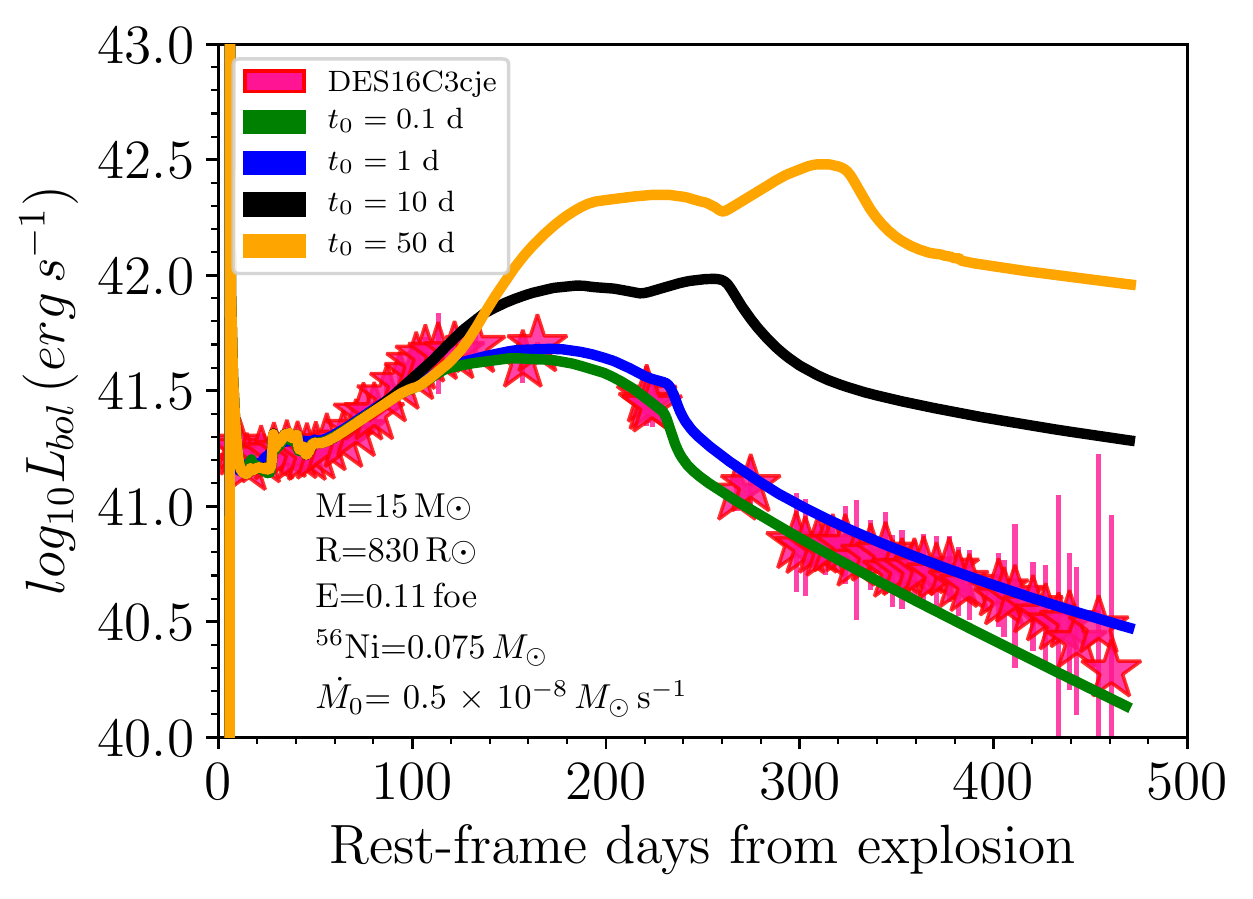}
\caption{Bolometric light curve of DES16C3cje (stars) compared with the results of the light curve calculations from hydrodynamic models. The continuous lines show the effect of $t_0$ in the 15~\Msun\ model. The used parameters are presented on the bottom.}
\label{fig:modelt0}
\end{figure*}
\clearpage

\setcounter{figure}{2}
\renewcommand\thefigure{\thesection.\arabic{figure}}  
\begin{figure*}
\centering
\includegraphics[width=0.85\textwidth]{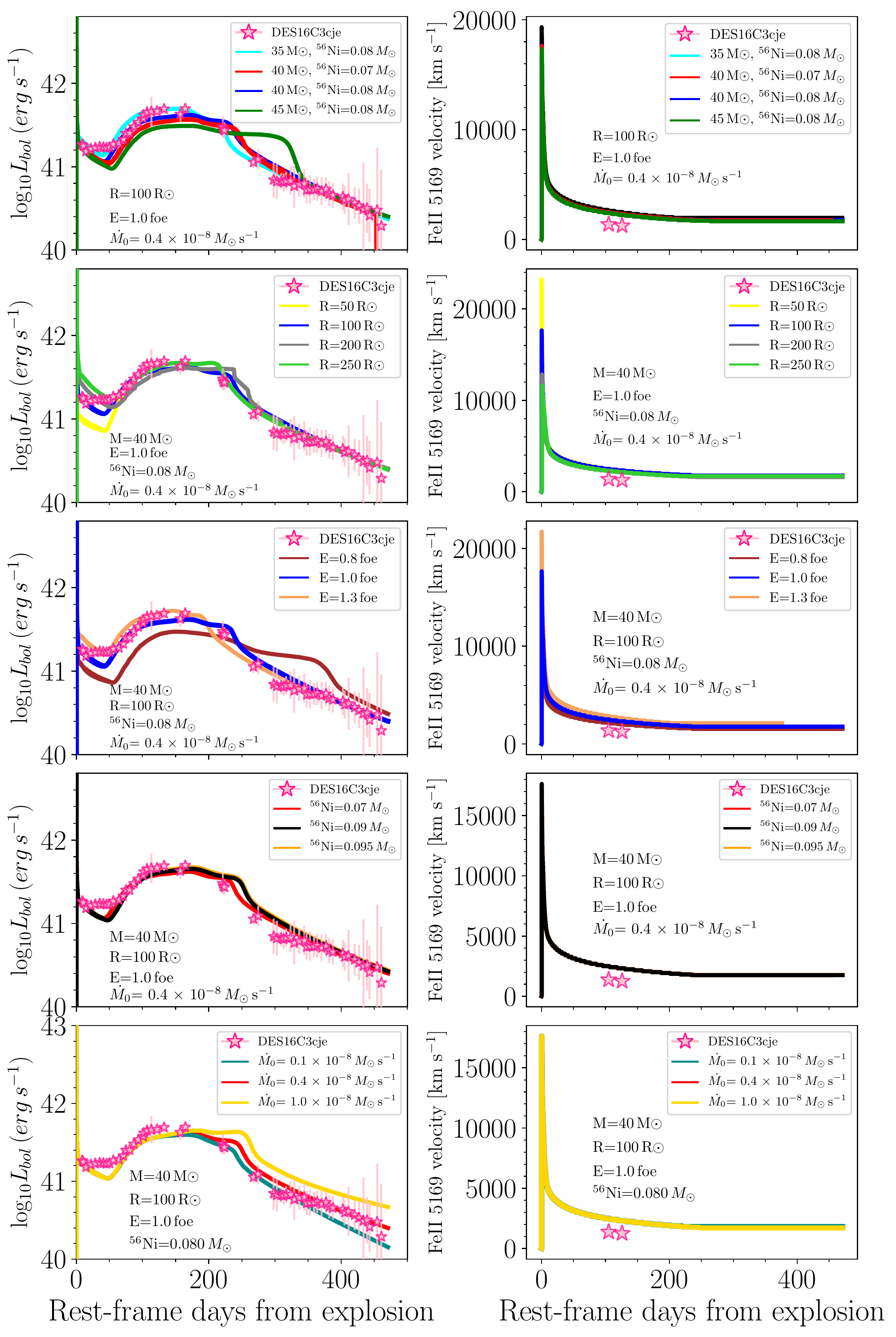}
\caption{Same as Figure~\ref{fig:model15} but for more massive and relatively compact progenitors.}
\label{fig:model40}
\end{figure*}
\clearpage

 \parbox{\textwidth}\par{
$^{1}$ Department of Physics and Astronomy, University of Southampton, Southampton, SO17 1BJ, UK\\
$^{2}$ Facultad de Ciencias Astron\'omicas y Geof\'isicas, Universidad Nacional de La Plata, Paseo del Bosque S/N, B1900FWA, La Plata, Argentina\\
$^{3}$ Instituto de Astrof\'isica de La Plata (IALP), CCT-CONICET-UNLP. Paseo del Bosque S/N, B1900FWA, La Plata, Argentina\\
$^{4}$ Kavli Institute for the Physics and Mathematics of the Universe (WPI), The University of Tokyo, 5-1-5 Kashiwanoha, Kashiwa, Chiba 277-8583, Japan\\
$^{5}$ School of Physics \& Astronomy, Cardiff University, Queens Buildings, The Parade, Cardiff, CF24 3AA, UK\\
$^{6}$ European Southern Observatory, Alonso de C\'ordova 3107, Casilla 19, Santiago, Chile\\
$^{7}$ Division of Science, National Astronomical Observatory of Japan, 2-21-1 Osawa, Mitaka, Tokyo 181-8588, Japan\\
$^{8}$ INAF -- Osservatorio Astronomico di Padova, Padova, Italy.\\
$^{9}$ Departamento de F\'isica Te\'orica y del Cosmos, Universidad de Granada, E-18071 Granada, Spain\\
$^{10}$ Lawrence Berkeley National Laboratory, 1 Cyclotron Road, Berkeley, CA 94720, USA\\
$^{11}$ University of Copenhagen, Dark Cosmology Centre, Juliane Maries Vej 30, 2100 Copenhagen O, Denmark\\
$^{12}$ The Oskar Klein Centre, Department of Astronomy, Stockholm University, AlbaNova, 10691 Stockholm, Sweden.\\
$^{13}$ INAF, Astrophysical Observatory of Turin, I-10025 Pino Torinese, Italy\\
$^{14}$ Max-Planck-Institut f{\"u}r Extraterrestrische Physik, Giessenbachstra\ss e 1, 85748, Garching, Germany\\
$^{15}$ School of Mathematics and Physics, University of Queensland,  Brisbane, QLD 4072, Australia \\
$^{16}$ Capodimonte Observatory, INAF-Naples , Salita Moiariello 16, 80131-Naples, Italy.\\
$^{17}$ European Southern Observatory, Karl-Schwarzschild-Str. 2, D-85748 Garching b. M\"unchen, Germany.\\
$^{18}$ Santa Cruz Institute for Particle Physics, Santa Cruz, CA 95064, USA\\
$^{19}$ School of Physics, O'Brien Centre for Science North, University College Dublin, Dublin, Ireland.\\
$^{20}$ Institute of Cosmology and Gravitation, University of Portsmouth, Portsmouth, PO1 3FX, UK\\
$^{21}$ CENTRA, Instituto Superior T\'ecnico, Universidade de Lisboa, Av. Rovisco Pais 1, 1049-001 Lisboa, Portugal\\
$^{22}$ Astronomical Observatory, University of Warsaw, Al. Ujazdowskie 4, 00-478 Warszawa, Poland\\
$^{23}$ Tuorla Observatory, Department of Physics and Astronomy, University of Turku, FI-20014 Turku, Finland\\
$^{24}$ Observatories of the Carnegie Institution for Science, 813 Santa Barbara Street, Pasadena, CA 91101, USA\\
$^{25}$ Sydney Institute for Astronomy, School of Physics, A28, The University of Sydney, NSW 2006, Australia\\
$^{26}$ School of Physics, Trinity College Dublin, The University of Dublin, Dublin 2, Ireland.\\
$^{27}$ Universit\'e Clermont Auvergne, CNRS/IN2P3, LPC, F-63000 Clermont-Ferrand, France\\
$^{28}$ Las Campanas Observatory, Carnegie Observatories, Casilla 601, La Serena, Chile\\
$^{29}$ Institute for Astronomy, University of Edinburgh, Royal Observatory, Blackford Hill, EH9 3HJ, UK \\
$^{30}$ Birmingham Institute for Gravitational Wave Astronomy and School of Physics and Astronomy, University of Birmingham, Birmingham B15 2TT, UK\\
$^{31}$ The Research School of Astronomy and Astrophysics, Australian National University, ACT 2601, Australia \\
$^{32}$ Departamento de F\'isica Matem\'atica, Instituto de F\'isica, Universidade de S\~ao Paulo, CP 66318, S\~ao Paulo, SP, 05314-970, Brazil\\
$^{33}$ Laborat\'orio Interinstitucional de e-Astronomia - LIneA, Rua Gal. Jos\'e Cristino 77, Rio de Janeiro, RJ - 20921-400, Brazil\\
$^{34}$ Fermi National Accelerator Laboratory, P. O. Box 500, Batavia, IL 60510, USA\\
$^{35}$ Instituto de Fisica Teorica UAM/CSIC, Universidad Autonoma de Madrid, 28049 Madrid, Spain\\
$^{36}$ CNRS, UMR 7095, Institut d'Astrophysique de Paris, F-75014, Paris, France\\
$^{37}$ Sorbonne Universit\'es, UPMC Univ Paris 06, UMR 7095, Institut d'Astrophysique de Paris, F-75014, Paris, France\\
$^{38}$ Department of Physics \& Astronomy, University College London, Gower Street, London, WC1E 6BT, UK\\
$^{39}$ Kavli Institute for Particle Astrophysics \& Cosmology, P. O. Box 2450, Stanford University, Stanford, CA 94305, USA\\
$^{40}$ SLAC National Accelerator Laboratory, Menlo Park, CA 94025, USA\\
$^{41}$ Centro de Investigaciones Energ\'eticas, Medioambientales y Tecnol\'ogicas (CIEMAT), Madrid, Spain\\
$^{42}$ Department of Astronomy, University of Illinois at Urbana-Champaign, 1002 W. Green Street, Urbana, IL 61801, USA\\
$^{43}$ National Center for Supercomputing Applications, 1205 West Clark St., Urbana, IL 61801, USA\\
$^{44}$ Institut de F\'{\i}sica d'Altes Energies (IFAE), The Barcelona Institute of Science and Technology, Campus UAB, 08193 Bellaterra (Barcelona) Spain\\
$^{45}$ INAF-Osservatorio Astronomico di Trieste, via G. B. Tiepolo 11, I-34143 Trieste, Italy\\
$^{46}$ Institute for Fundamental Physics of the Universe, Via Beirut 2, 34014 Trieste, Italy\\
$^{47}$ Observat\'orio Nacional, Rua Gal. Jos\'e Cristino 77, Rio de Janeiro, RJ - 20921-400, Brazil\\
$^{48}$ Department of Physics, IIT Hyderabad, Kandi, Telangana 502285, India\\
$^{49}$ Department of Astronomy/Steward Observatory, University of Arizona, 933 North Cherry Avenue, Tucson, AZ 85721-0065, USA\\
$^{50}$ Jet Propulsion Laboratory, California Institute of Technology, 4800 Oak Grove Dr., Pasadena, CA 91109, USA\\
$^{51}$ Institut d'Estudis Espacials de Catalunya (IEEC), 08034 Barcelona, Spain\\
$^{52}$ Institute of Space Sciences (ICE, CSIC),  Campus UAB, Carrer de Can Magrans, s/n,  08193 Barcelona, Spain\\
$^{53}$ Kavli Institute for Cosmological Physics, University of Chicago, Chicago, IL 60637, USA\\
$^{54}$ Department of Astronomy, University of Michigan, Ann Arbor, MI 48109, USA\\
$^{55}$ Department of Physics, University of Michigan, Ann Arbor, MI 48109, USA\\
$^{56}$ Department of Physics, Stanford University, 382 Via Pueblo Mall, Stanford, CA 94305, USA\\
$^{57}$ Center for Cosmology and Astro-Particle Physics, The Ohio State University, Columbus, OH 43210, USA\\
$^{58}$ Department of Physics, The Ohio State University, Columbus, OH 43210, USA\\
$^{59}$ Center for Astrophysics $\vert$ Harvard \& Smithsonian, 60 Garden Street, Cambridge, MA 02138, USA\\
$^{60}$ Australian Astronomical Optics, Macquarie University, North Ryde, NSW 2113, Australia\\
$^{61}$ Lowell Observatory, 1400 Mars Hill Rd, Flagstaff, AZ 86001, USA\\
$^{62}$ Department of Physics and Astronomy, University of Pennsylvania, Philadelphia, PA 19104, USA \\
$^{63}$ Instituci\'o Catalana de Recerca i Estudis Avan\c{c}ats, E-08010 Barcelona, Spain\\
$^{64}$ Department of Astrophysical Sciences, Princeton University, Peyton Hall, Princeton, NJ 08544, USA\\
$^{65}$ Brandeis University, Physics Department, 415 South Street, Waltham MA 02453\\
$^{66}$ Computer Science and Mathematics Division, Oak Ridge National Laboratory, Oak Ridge, TN 37831\\
$^{67}$ Universit\"ats-Sternwarte, Fakult\"at f\"ur Physik, Ludwig-Maximilians Universit\"at M\"unchen, Scheinerstr. 1, 81679 M\"unchen, Germany\\
$^{68}$ Cerro Tololo Inter-American Observatory, National Optical Astronomy Observatory, Casilla 603, La Serena, Chile\\
$^{69}$ Department of Physics and Astronomy, Pevensey Building, University of Sussex, Brighton, BN1 9QH, UK\\
}


\bsp	
\label{lastpage}
\end{document}